\documentclass[%
12pt,
 oneocolumn,
 nofootinbib,
 amsmath,amssymb,
 aps,
floatfix,
unsortedaddress
]{revtex4}

\usepackage{graphicx,color}
\usepackage{subfig}
\usepackage{dcolumn}% Align table columns on decimal point
\usepackage{bm}% bold math
\usepackage{graphicx,color,caption}
\usepackage[font=small,justification=Justified,format=plain]{caption}
\usepackage{float}

\usepackage[pdfpagelabels, pdfencoding=auto, psdextra]{hyperref}
\hypersetup{%
 pdfsubject=Paper,
 pdfkeywords={nuclear physics} {Pionless EFT} {Two-Nucleon},
 unicode = true,
 breaklinks = true,
 colorlinks = true,
 linkcolor = blue,
 citecolor = blue,
 menucolor = blue,
 citecolor = blue,
 urlcolor = blue
}

\graphicspath{{/}}

\begin{document}

\title{Low--Energy Deuteron--Alpha Elastic Scattering in Cluster Effective Field Theory}

\author{F. Nazari}
 \email{f.nazari@email.kntu.ac.ir}
 \author{M. Radin}
\email{radin@kntu.ac.ir}
%\thanks{(Corresponding author)}
 \affiliation{Department of Physics, K. N. Toosi University of Technology, P.O.Box 16315-1618, Tehran, Iran
}

\author{M. Moeini Arani }
\email{m.moeini.a@ut.ac.ir}
\affiliation{Malek Ashtar University of Technology, Tehran, Iran}

\begin{abstract}
In this paper, we study the low-energy $d-\alpha$ elastic scattering
within the two-body cluster effective field theory (EFT) framework.
The importance of the $d(\alpha,\alpha) d$ scattering in the $^6 \textrm{Li} $
production reaction leads us to study this system in an effective
way. In the beginning, the scattering amplitudes of each channel are
written in a cluster EFT with two-body formalism. Using the
effective range expansion analysis for the elastic scattering phase
shift of $S$, $P$ and $D$ partial waves, the unknown EFT low-energy coupling
constants are determined and the leading and next-to-leading orders
EFT results for the phase shift in each channel are presented. To
verify the accuracy of the results, we compare experimental phase
shift and differential cross section data with obtained results. The
accuracy of the EFT results and consistency with the experimental
data indicate that the EFT is an effective approach for describing
low-energy systems.
   
\end{abstract}
\maketitle

%%%%%%%%%%%%%%%%%
\section{Introduction\label{sec:1}}

The  $d- \alpha$ elastic scattering has been of interest for many
years as a source of information about the low-lying $T\!=\!0$ states of
$^{6}\textrm{Li}$. The analysis of  $d-\alpha$ elastic
scattering data, to obtain the correct
energy dependent phase shifts of this process and determine the
corresponding level parameters of the $^{6}\textrm{Li}$ nucleus, has
been studied widely in the past decades. The $d-\alpha$ scattering
has been studied extensively in the past\cite {L. C. Mc lntyre, A. Jenny,P. Marmier, W. Gr,M. Bruno,I. Koersner,I. Slaus,P. Niessen,Y. Koike,K. Hahn, A. Galonsky}, and the low-lying levels of
$^{6}\textrm{Li}$ have been extensively investigated both
experimentally and theoretically \cite{L. S. Senhouse, K. W. Allen, W. C. Barber, T. A. Romanowski,D. R. Inglis, P. H. Wackman}. Recently, the $d-\alpha$ scattering was investigated using the screening and
renormalization approach in the framework of momentum space
three-particle equations \cite{A. Deltuva}.

In the present work, we focus on applying the effective field theory
(EFT) formalism as a model-independent, systematic and
controlled-precision procedure for the investigation of
$d-\alpha$ elastic scattering at the center-of-mass (CM) energies
about a few MeV corresponding to the validity of the EFT expansion. The
applications of EFT approach in the few-nucleon systems have been widely
studied \cite{Bedaque-van Kolck, Braaten-Hammer, Kaplan-S-W, Phillips-R-S, Chen-R-S}. Also, in
recent years the nuclear systems with $A\!>\!4$ which can be classified
in the two-body sector are studied by halo EFT scheme~\cite{Hammer-Phillips}. The deuteron can be thought of as the
simplest halo nucleus whose core is a nucleon, however, there are
some EFT works that the deuteron field is introduced as an
elementary-like field
\cite{Ando-nd, Ando-Yang-Oh, MMA-dt-na, MMA-dt-dt}. Halo EFT captures
the physics of resonantly $P$-wave interactions in $n -\alpha$
scattering up to next-to-leading order (NLO)~\cite{bertulani-et
al,Bedaque-et al} and studying two-neutron halo system
$^6$$\textrm{He}$~\cite{chen-ji,MMA-nna}. The effects of the Coulomb
interaction in two-body systems such as $p-$$p$
\cite {33, Kong-Ravndal01, Barford-Birse, Ando-S-H-H, Ando-Birse},
$p-$$^7$$\textrm{Li}$~\cite{Lensky-Birse},
$\alpha-$$^{12}\textrm{C}$ \cite{Ando-12C}, and
$\alpha-\alpha$ scattering \cite{Higa-Hammer} and
$^3$$\textrm{He}(\alpha,\gamma)^7$$\textrm{Be}$
\cite{Higa-Rupak-Vaghani}, have been considered by the EFT approach.

Before applying the EFT method to the description of low-energy
$d-\alpha$ radiative capture, we construct the
EFT formalism for the $d-\alpha$ scattering in the current study. Although $d-\alpha$
is a six-nucleon system, at low energies, to a good approximation,
the alpha particle may be considered a spin zero structureless
boson, and thereby the theoretical description of $d-\alpha$
scattering may be reduced to a three-body problem made up of one
alpha and two nucleons. At the low-energy regime below deuteron
breakup, we can take into account that both deuteron and alpha nucleus as
point-like and structureless particles. Therefore, our present EFT
for low-energy $d-\alpha$ scattering is constructed using the
two-body cluster formalism. The phase shift analysis and
differential cross section calculation for the elastic $d-\alpha$
scattering procedure, after determination of the unknown EFT
low-energy coupling constants (LECs), are the main purposes of this
paper. We obtain the EFT LECs by using available low-energy
experimental data for the elastic $d-\alpha$ scattering. Here, we
study the scattering into the $S$-, $P$- and $D$-wave states using
the effects corresponding to the scattering length, effective range
 and shape parameter at each channel. The evaluated results can help
us to investigate the astrophysical radiative capture processes
$d+\alpha\rightarrow ^6$$\textrm{Li}+\gamma$ using halo/cluster EFT formalism in the future.

The manuscript is organized as follows. In Sec.~\ref{sec:2},
 the pure Coulomb and Coulomb-subtracted amplitudes of the
$d-\alpha$ scattering in all possible $l\!=\!0,~1,~2$ partial waves
using the effective range expansion (ERE) and EFT formalisms are calculated. The
values of the unknown EFT LECs are determined by matching our
relations of phase shift to the available low-energy experimental
data in Sec.~\ref{sec:3}. Using the power counting analysis of the
effective range parameters, we plot the EFT differential cross
section against CM energy and angle with the dominant scattering
amplitudes and compare with the available
data in Sec.~\ref{sec:4}. We summarize the paper and discuss extension of the
investigation to other few-body systems in Sec.~\ref{sec:5}.

\section{Scattering amplitude}{\label{sec:2}}
In this section, the pure Coulomb and Coulomb-subtracted
scattering amplitudes for the two-body $d - \alpha$ elastic scattering
using cluster EFT formalism are extracted. The elastic scattering amplitude for
two particles  interacting via short-range strong and long-range
Coulomb interactions in the CM framework is written as
\begin{equation}\label{eq:1}
T(\textbf{p}',\textbf{p};E)=T_C(\textbf{p}',\textbf{p};E)+T_{CS}(\textbf{p}',\textbf{p};E),
\end{equation}
where $T_{C}$ indicates the pure Coulomb scattering amplitude and
$T_{CS}$ represents the scattering amplitude for the strong
interaction in the presence of the Coulomb interaction with
$E\!=\!\frac{p^2}{2\mu}$ as the CM energy of the system. $\textbf{p}$
and $\textbf{p}'$ denote the relative momentum of incoming and
outgoing particles, respectively~\cite{Higa-Hammer}.

\subsection{Pure Coulomb amplitude}{\label{subsec:1}}.

\begin{figure}
 \centering
\includegraphics[width=5.5in,height=6.2cm]{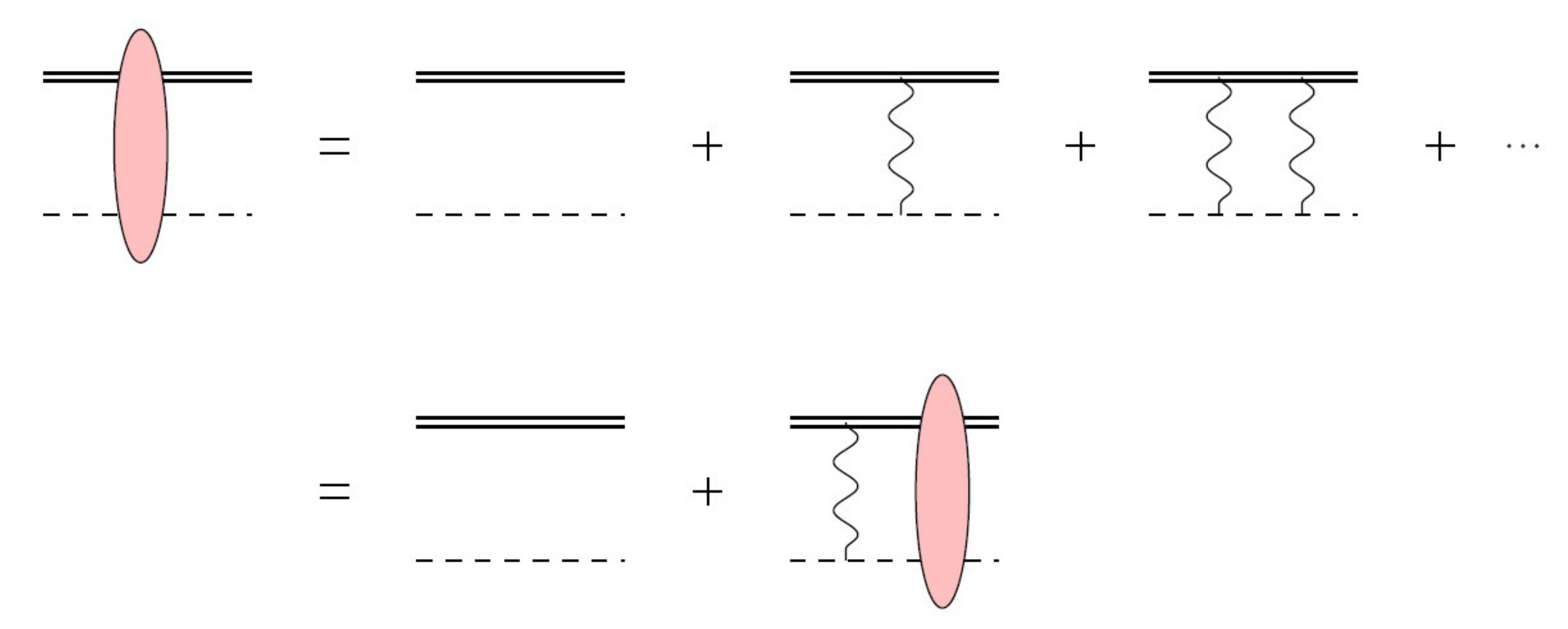}
\caption{\small{Coulomb ladder diagrams. The single dashed and double lines represent the scalar $\alpha$ and vector deuteron particle, respectively. The wavy lines represent the exchanged photons}.}\label{fig:1}
\end{figure}

The strength of the Coulomb-photon exchanges is provided by the
dimensionless Sommerfeld parameter which for the $d-\alpha$
interaction can be written as
\begin{equation}\label{eq:2}
\eta_p=\frac{k_C}{p}=\frac{Z_{\alpha}Z_d\,\alpha_{em}\,\mu}{p}.
\end{equation}
Here $k_C$ is the inverse of the Bohr radius of the $d- \alpha$
system, $\alpha_{em}\!\equiv\! e^2/4\pi \!\sim \!1/137$ represents the fine
structure constant, $p$ is the relative momentum of two particles in
the CM framework, $ Z_{\alpha} (Z_d)$ indicates the atomic numbers
of alpha (deuteron), and $\mu$ denotes the reduced mass of
$d-\alpha$ system. Based on the fact that each photon-exchange
insertion is proportional to $\eta_p$ so, in the low-energy
scattering region, $p\lesssim k_C$, we should consider the full
Coulomb interaction non-perturbatively as depicted in
Fig.~\ref{fig:1}. In order to consider the Coulomb contribution in
the two-body  $d-\alpha$ system, we use the Coulomb Green's function
as follows. According to Fig.~\ref{fig:1}, the Coulomb Green's
function is related to the free Green's function through the
integral equation as \cite{goldberger1964collision}
\begin{equation}\label{eq:3}
\hat{G}_C^{\pm}=\hat{G}_0^{\pm}+\hat{G}_0^{\pm}\,\hat{V}_C\,\hat{G}_C^{\pm},
\end{equation}
where the free and Coulomb Green's functions for the $d-\alpha$
system are given by
\begin{equation}\label{eq:4}
\hat{G}_0^{\pm}=\frac{1}{E-\hat{H}_0\pm i\epsilon},\quad\quad \hat{G}_C^{\pm}=\frac{1}{E-\hat{H}_0-\hat{V}_C\pm i\epsilon},
\end{equation}
with $\hat{V}_C\!=\!2\alpha_{em}/r$ and
$\hat{H}_0\!=\!\frac{\hat{p}^2}{2\mu}$ as the repulsive Coulomb
potential between alpha and deuteron and the free-particle
Hamiltonian, respectively. The signs $(\pm)$ are corresponding to the
retarded and advanced Green's functions.
The incoming and outgoing Coulomb
wave functions can be obtained by solving the Schrodinger
equation with the full Hamiltonian $\hat{H}\! = \!\hat{H}_0\! + \!\hat{V}_C$ as~\cite{33,holstein1999hadronic}
\begin{eqnarray}\label{eq:5}
\chi_p^{(\pm)}(\textbf{r})=\sum _{l=0}^{\infty}(2l+1)i^l e^{i\sigma_l}P_l(\hat {\textbf{p}}\cdot\hat{\textbf{r}})\, C_l(\eta_p)\,(rp)^l e^{\mp i\textbf{p}\cdot\textbf{r}}\textrm{M}(l+1\pm i\eta_p,2l+2;\pm 2ipr),\quad
\end{eqnarray}
where $\textrm{M}(a,b;z)$ is well-known Kummer function, $P_l$ denotes the Legendre function and
$\sigma_l\!=\!\arg \mathrm{\Gamma}(l+1+i\eta_p)$ indicates the pure Coulomb phase
shift~\cite{abromowitz1965handbook}. The normalized
constant $C_l(\eta_p)$ is always positive and has the form
\begin{eqnarray}\label{eq:6}
C^2_l(\eta_p)=\frac{2^{2l}C_0^2(\eta_p)\,\prod_{n=1}^l(n^2+\eta_p^2)}{\mathrm{\Gamma}(2l+2)^2},
\end{eqnarray}
where $C_0^2(\eta_p)$, the probability to find the two interacting particles at
zero separation, is defined as
\begin{eqnarray}\label{eq:7}
C_0^2(\eta_p)=\chi_{p'}^{(\pm)}(\mathbf{0})\chi_{p}^{*(\pm)}(\mathbf{0})=\frac{2\pi \eta_p}{e^{2\pi \eta_p}-1}.
\end{eqnarray}
According to the expression of the Coulomb wave function of
Eq.~(\ref{eq:5}), the partial wave expansion of the pure Coulomb
amplitude is given by \cite{gaspard2018connection}
\begin{eqnarray}\label{eq:8}
T_C(\textbf{p}',\textbf{p};E)&=&\langle
\mathbf{p}'|\hat{V}_C|\chi_{p}^{(+)}\rangle= \sum_{l=0}^{\infty}
(2l+1) T_C^{[l]}
P_l(\hat{\mathbf{p}}'\cdot\hat{\mathbf{p}})\nonumber\\
&=&-\frac{2
\pi}{\mu} \sum_{l=0}^{\infty} (2l+1)
\frac{e^{2i\sigma_l}-1}{2ip}P_l(\hat{\mathbf{p}}'\cdot\hat{\mathbf{p}})\nonumber\\
&=&\frac{2
\pi}{\mu}\frac{\eta_p^2}{2k_C} \,\csc^2(\theta/2) \,
\textrm{exp} \big[2i\sigma_0-2i \eta_p  \,\ln
(\sin(\theta/2))\big],\qquad
\end{eqnarray}
where $\cos \theta=\hat{\mathbf{p}}'\cdot\hat{\mathbf{p}}$ and
$p=|\mathbf{p}|=|\mathbf{p}'|$. This is the well-known Mott
scattering amplitude which holds at very low energies \cite{bethe1949theory}.

\subsection{Coulomb-subtracted scattering amplitude}{\label{subsec:2}}
The strong scattering amplitude modified by the Coulomb corrections is
\begin{eqnarray}\label{eq:9}
% \nonumber to remove numbering (before each equation)
T_{CS}(\mathbf{p}',\mathbf{p};E)=\langle\chi_{p'}^{(-)}|\hat{V}_S|\mathrm{\Psi}_{p}^{(+)}\rangle,
\end{eqnarray}
where $|\mathrm{\Psi}_{p}^{(+)}\rangle$ represent incoming state for Coulomb-distorted short-range interaction, while $\hat{V}_S$ is the short-range interaction operator. The amplitude $T_{CS}$ can be expressed in the partial wave decomposition as \cite{33}
\begin{equation}\label{eq:10}
T_{CS}(\mathbf{p}',\mathbf{p};E)=\sum_{l=0}^{\infty}(2l+1)T^{[l]}_{CS}(p)\,e^{2i \sigma_l}P_l(\mathbf{p}'\cdot \mathbf{p}),
\end{equation}
with
\begin{eqnarray}\label{eq:11}
T^{[l]}_{CS}(p)&=&-\frac{2\pi}{\mu}\frac{1}{p (\textrm{cot}\delta_l-i)},\quad\quad\,\,\,
\end{eqnarray}
where $\delta_l$ denotes the Coulomb-corrected phase shift. The
Coulomb-subtracted amplitude $T^{[l]}_{CS}$ can usually be expressed
in terms of a modified ERE as \cite{Ando-12C}
\begin{eqnarray}\label{eq:12}
T^{[l]}_{CS}(p) &=&-\frac{2\pi}{\mu}\frac{C_0^2(\eta_p)\,W_l(\eta_p)}{K_l(p)-H_l(\eta_p)},\quad
\end{eqnarray}
with
\begin{eqnarray}\label{eq:13}
W_l(\eta_p) &=&\frac{k_C^{2l}}{(l!)^2}\,\prod^l_{n=0}(1+\frac{n^2}{\eta_p^2}),
\\
H_l(\eta_p)&=&2k_C W_l(\eta_p)H(\eta_p),\\
H(\eta_p)&=&\psi(i\eta_p)+\frac{1}{2i\eta_p}-\ln(i\eta_p),
\end{eqnarray}
where the function $\psi$ is the logarithmic derivative of
 Gamma function. The function $K_l(p)$ represents the interaction
due to the short-range strong interaction which is obtained in terms
of the effective range parameters as \cite{bethe1949theory}
\begin{eqnarray}\label{eq:16}
% \nonumber to remove numbering (before each equation)
K_l(p)&=&-\frac{1}{a_l}+\frac{1}{2}r_l\,p^2+\frac{1}{4}s_l\,p^4+\cdots,
\end{eqnarray}
with $a_l$, $r_l$ and $s_l$ as the scattering length, effective range and shape parameter, respectively.

\subsection{Scattering amplitudes in cluster EFT approach}{\label{subsec:3}}
In the present study, we consider the deuteron and alpha as the
point-like particles, so the degrees of freedom of the $d-\alpha$
system in the current cluster EFT are only alpha  and deuteron.
At the low-energy regime, the $S$, $P$ and $D$ partial waves have the
dominant contributions in the $d-\alpha$  elastic scattering
amplitude. We should point out that the available low-energy
experimental data for the differential cross section of the elastic
$d-\alpha$ scattering show a resonance below the CM energy 1 MeV.
Theoretically, this resonance can be constructed only by including
the $D$-wave effects in the cross section. Also, the dominant
contribution of the deuteron radiative capture by alpha particles at
energy above 0.5 MeV comes from E2 transition with incoming $D$-wave
states \cite {Higa-Hammer,Andophys2006,braun2019electric}. Therefore, we consider the
$D$-wave scattering amplitudes of the $d-\alpha$ system in the
present low-energy study. So, according to the spin zero of alpha
and spin one of the deuteron and considering the $l$-wave components
of the $d-\alpha$ system, the possible states for the two-body
$d-\alpha$ system are $\xi \equiv$ $^3\!S_1$, $^3\!P_0$, $^3\!P_1$,
$^3\!P_2$, $^3\!D_1$, $^3\!D_2$ and $^3\!D_3$ corresponding to the total
angular momentums, $J = 0, 1, 2, 3$.

At the low-energy regime, $p\! \leq \! k_C \sim 18~ \textrm{MeV}$, the
on-shell CM momentum of the system is scaled as low-momentum $Q$.
The high-momentum scale is set by the lowest energy degrees of
freedom that has been integrated out. According to the fact that
there is no existing explicit pions and any deuteron deformation,
the high-momentum scale $\mathrm{\Lambda}$ has been chosen between the pion
mass, $m_\pi\sim 140$ MeV and the momentum corresponding to the
deuteron binding energy, $B_d$ i.e., $\sqrt{2m_d B_d} \sim 90$ MeV.
Around the $p\sim k_C\sim18$ MeV, the expansion parameter of the
current EFT is estimated of the order $1/5$. Increasing the energy,
the expansion deteriorates and the precision of our EFT prediction will
be questionable for $E_{CM}=\frac{p^2}{2\mu}>3.3$ MeV. The
Sommerfeld parameter $\eta_p$ is enhanced by decreasing the energy.
So, $\eta_p$ would be large around $p\!\lesssim\! k_C$ and the elastic
scattering amplitude requires non-perturbative treatment of the
Coulomb photons.

The non-relativistic Lagrangian for the strong interactions in the
$d-\alpha$ system involving the invariance under small-velocity
Lorentz, parity and time-reversal transformations and describing the
dynamics in all feasible channels is given by
\begin{eqnarray}\label{eq:17}
\mathcal{L}^{[\xi]}&=&\phi^{\dagger}(i\partial _0+\frac{\nabla ^2}{2m_\alpha})\phi+d_i^{\dagger}(i\partial _0+\frac{\nabla ^2}{2m_d})d_i +\,\eta^{[\xi]}t^{[\xi]^\dagger}\!\Big[i\partial _0\!+\!\frac{\nabla ^2}{2m_t}-\mathrm{\Delta}^{[\xi]}\Big]\!t^{[\xi]}\nonumber\\
&&+\,g^{[\xi]}[t^{[\xi]^\dagger}\!(\phi\, \mathrm{\Pi}^{[\xi]}d)+h.c.] +\,h^{[\xi]}t^{[\xi]^\dagger}\!\Big[(i\partial _0\!+\!\frac{\nabla ^2}{2m_t})^2\Big]t^{[\xi]}
+\cdots,
\end{eqnarray}
where "$\cdots$" stands for the terms with more derivatives and/or
auxiliary fields. The scalar field $\phi$ represents the spinless
$\alpha$ field with mass $m_\phi\! =\! 3727.38 ~\textrm{MeV}$ and the
vector field $d_i\!=\!\varepsilon_i^d d$ indicates the deuteron nucleus
axillary field with mass $m_d \!=\!1875.61 ~\textrm{MeV}$. The sign
$\eta^{[\xi]}$ is used to match the sign of the effective range
$r^{[\xi]}$ and reflects the auxiliary character of the dimeron
field. The dimeron field $t^{[\xi]}$ with mass $m_t \!=\! m_d \!+\!
m_{\phi}$, and ${\mathrm{\Pi}}^{[\xi]}$ operator for each $\xi$ channel are
defined as
\begin{eqnarray}\label{eq:18}
t^{[\xi]}&=&\left\lbrace \begin{array}{lc}
    \bar{t}_i,  &\qquad\qquad\quad \, \xi={^3\!S_1}\\
    t, & \,\qquad\qquad\quad \xi={^3\!P_0}\\
    t_{k}, &\,\qquad\qquad\quad \xi={^3\!P_1} \\
    t_{ij}, & \,\qquad\qquad\quad \xi={^3\!P_2}\\
    \tilde{t}_j, & \,\qquad\qquad\quad \xi={^3\!D_1}\\
    \tilde{t}_{kl}, & \,\qquad\qquad\quad \xi={^3\!D_2}\\
    \tilde{t}_{kji}, &\,\qquad\qquad\quad \xi={^3\!D_3}
    \!\!\end{array}\right\rbrace,\,\qquad
\\
\mathrm{\Pi}^{{[\xi]}}&=&\left\lbrace \begin{array}{lc}
    \varepsilon_i^d,  & \,\xi={^3\!S_1}\\
    \sqrt{3}\,\mathcal{P}_i\,\varepsilon_i^d, & \, \xi={^3\!P_0}\\
    \sqrt{3/2}\,\epsilon_{kji}\,\mathcal{P}_j\,\varepsilon_i^d ,&\, \xi={^3\!P_1} \\
    3/\sqrt{5}\,\mathcal{P}_j\,\varepsilon_i^d, & \, \xi={^3\!P_2}\\
    3/\sqrt{2}\,\tau_{ji}\,\varepsilon_i^d, & \, \xi={^3\!D_1}\\
    \sqrt{3/2}\,\epsilon_{ijl}\,\tau_{kj}\,\varepsilon_i^d,  & \, \xi={^3\!D_2}\\
    \sqrt{45/8}\,\tau_{kj}\,\varepsilon_i^d, & \, \xi={^3\!D_3}
    \end{array}\right\rbrace,\,\,\,\qquad
\end{eqnarray}
where the derivative operators are introduced as
\begin{eqnarray}\label{eq:19}
\mathcal{P}_i=\frac{1}{i}(\frac{\mu}{\overrightarrow{m}} \overrightarrow{\nabla_i}
 -\frac{\mu}{\overleftarrow{m}} \overleftarrow{\nabla_i} ),\quad\,
\tau_{ij}
 =\mathcal{P}_i \mathcal{P}_j -\frac{1}{3}\delta_{ij}\mathcal{P}_k\mathcal{P}_k.\quad
\end{eqnarray}
In the following, the coupling constants $\mathrm{\Delta}^{[\xi]}$,
$g^{[\xi]}$, and $h^{[\xi]}$ for channel $\xi$ are related to the
corresponding scattering length, effective range and shape
parameter.

\begin{figure}
\begin{center}
\includegraphics[width=5.3in,height=5.5cm]{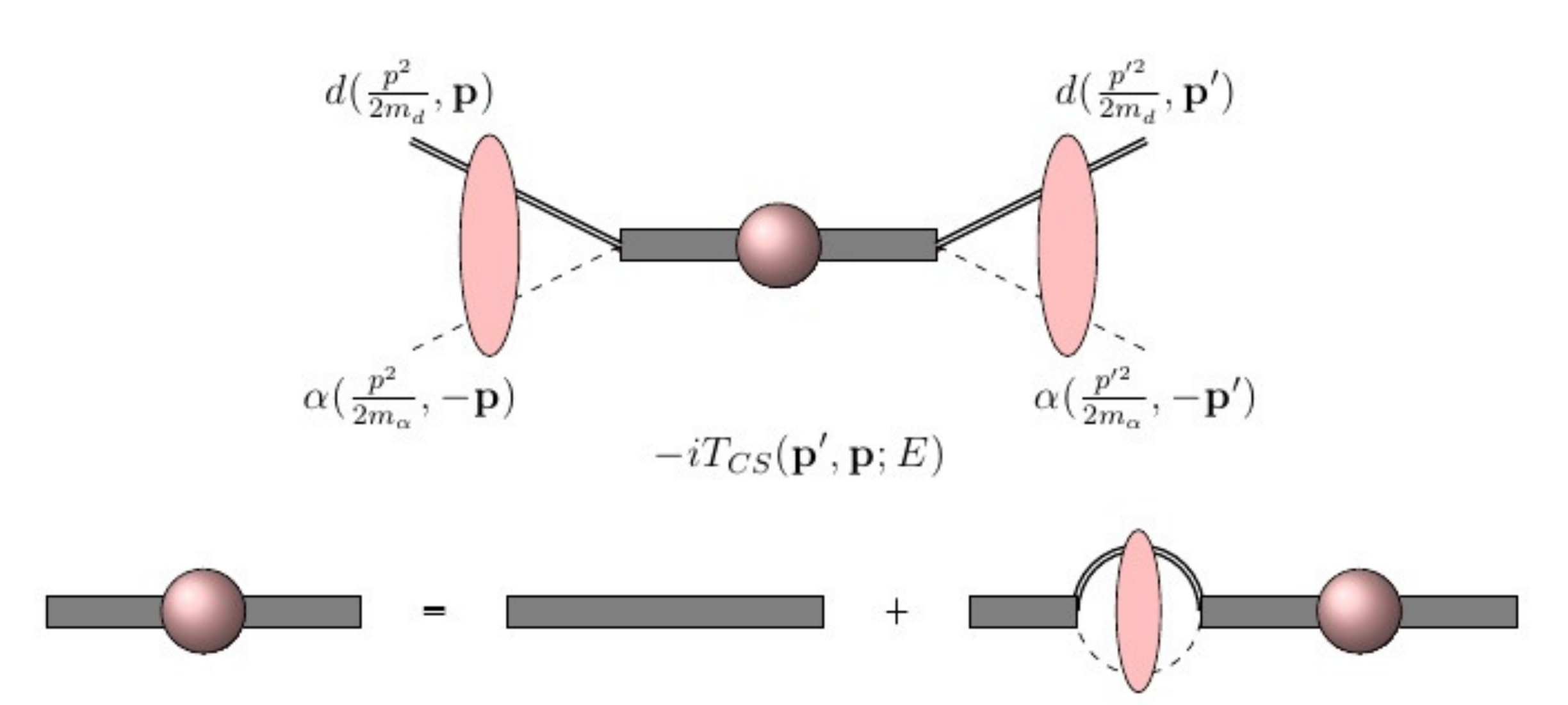}
\end{center}
\captionof{figure}{\small{The amplitude of the $d-\alpha$ elastic scattering.
The thick line is the bare dimeron propagator and the thick dashed line with a filled circle represents the full dimeron
propagator. All remained notations are the same as in Fig.~\ref{fig:1}}.}\label{fig:2}
\end{figure}

The cluster EFT diagram of the $d-\alpha$ elastic scattering
amplitude is shown in Fig. \ref{fig:2}. According to this diagram the
building block of the  scattering amplitude is the full propagator
of the dimeron. The bare and full propagators used in
$T_{CS}^{[\xi]}(\mathbf{p}',\mathbf{p},E)$ are depicted by the thick  line and the
thick line with filled circle, respectively. To
evaluate the EFT results for the $d-\alpha$ elastic scattering
amplitude in channel $\xi$, the external legs should be attached to
the full dimeron propagator as shown in the first line of Fig.~\ref{fig:2}. So, the Coulomb-subtracted EFT amplitudes of the
on-shell $d-\alpha$ scattering for each channel $\xi$ can be
evaluated by
\begin{eqnarray}\label{eq:20}
-i(2l+1)T_{CS}^{[\xi]}(p)P_l(\hat{\textbf{p}}'\cdot \hat{\textbf{p}}) e^{2i\sigma_l}= -ig^{[\xi]^2}D^{[\xi]}(E,\textbf{0}) C^2_0(\eta_p)W_l(\eta_p) P_l(\hat{\textbf{p}}'\cdot \hat{\textbf{p}}) e^{2i\sigma_l}\quad\quad.
\end{eqnarray}
The detailed derivations of Eq.~(\ref{eq:20}) for all channels are
presented in Appendix A.
Here, without any estimation for the values of effective range
parameters, we introduce the initial scheme in which the LO
contribution of Coulomb-subtracted $d-\alpha$ scattering for
channels $\xi \equiv$ $^3\!S_1$, $^3\!P_0$, $^3\!P_1$, and $^3\!P_2$ are
calculated using the first four terms in Lagrangian (\ref{eq:17})
and the last term initially enters as NLO corrections as in some
literature on the halo/cluster EFT \cite {Hammer-Phillips,rupak2016radiative,ryberg2014effective}. However, the properties
of the $D$-wave states are somewhat different. For the LO
calculation of the $D$ waves, taking into account $\eta^{[\xi]}\!=\!\pm
1$, we should include three EFT LECs in our Lagrangian
(\ref{eq:17}), namely, $\mathrm{\Delta}^{[\xi]}$, $g^{[\xi]}$, and $h^{[\xi]}$
corresponding to the scattering length, effective range, and shape
parameter. The additional second-order kinetic term constant
$h^{[\xi]}$ is needed to renormalize the interacting $D$-wave
propagator which contains up to quintic divergences \cite{braun2019electric}.
According to this suggested scheme, the LO contribution of the
scattering amplitude in channels $\xi={^3\!S_1},$ $ {^3\!P_0}$, ${^3\!P_1}
$, and ${^3\!P_2}$ is constructed by both their scattering lengths and
effective ranges and their shape parameter influences are
considered as NLO correction. However, for
$\xi={^3\!D_1},$ ${^3\!D_2} $, and ${^3\!D_3}$ channels, all the scattering
lengths, effective ranges, and shape parameters insert in the
scattering amplitude at LO.

So, with respect to Fig.~\ref{fig:2}, up-to-NLO full dimeron
propagator for $l\!=\!0$ and 1 channels in the CM framework can be
evaluated by
\begin{eqnarray}\label{eq:21}
D^{[\xi]}(E,\textbf{0})&=&\frac{\eta^{[\xi]}}{E\!-\!\mathrm{\Delta}^{[\xi]}
-\!\frac{1}{2l+1}\eta^{[\xi]}g^{[\xi]^2}J_l(E)}
 \Big[\underbrace{\!\!\!1\!\!\!\!_{_{_{_{_{_{_{_{_{_{_{}}}}}}}}}}}}_{\mathrm{LO}}
\!\!-\underbrace{\frac{\eta^{[\xi]}h^{[\xi]}E^2}{E\!-\!\mathrm{\Delta}^{[\xi]}
-\!\frac{1}{2l+1}\eta^{[\xi]}g^{[\xi]^2}J_l(E)}}_{\mathrm{NLO ~corection}}\Big]\!.\quad\quad\,
\end{eqnarray}
and taking into consideration the suggested scheme for the channels
$\xi \equiv$ $^3\!D_1$, $^3\!D_2$, and $^3\!D_3$ all terms in
Eq.~(\ref{eq:16}) should be considered at LO and so, the full
dimeron propagator for these channels is obtained by
\begin{equation}\label{eq:22}
D^{[\xi]}(E,\textbf{0})\!=\!\frac{\eta^{[\xi]}}{E\!-\!\mathrm{\Delta}^{[\xi]}\!+\!h^{[\xi]}E^2-\!\frac{1}{2l+1}\eta^{[\xi]}g^{[\xi]^2}J_l(E)}.
\end{equation}
The fully dressed bubble $J_{l}$ in Eqs.~(\ref{eq:21}) and
(\ref{eq:22}), which is described for the propagation of the
particles from initially zero separation and back to zero separation
for each channel, is divergent and should be regularized. We regularize the divergence by dividing the integral $J_l$ into two finite and infinite parts as $J_l\! = \!J_l^{fin}
\! + \!J_l^{div}$ \cite{kamand2013effective} . The detailed of this regularization for all channels are presented in Appendix A.
The finite part is obtained as \cite {ando2007low}
\begin{eqnarray}\label{eq:23}
J^{fin}_l(p)=-\frac{\mu}{2\pi}H_l(\eta_p).
\end{eqnarray}
The divergent part is momentum-independent for the $S$-wave and are sum up momentum-independent and momentum squared
parts for the
$P$-waves. For the $D$-waves, the divergences are divided into three
parts, momentum-independent, momentum-squared and momentum-cubed.
These divergences absorbed in $\mathrm{\Delta}^{[\xi]}$, $g^{[\xi]}$ and $h^{[\xi]}$
parameters via introducing the renormalized parameters $\mathrm{\Delta}_{\!R}^{[\xi]}$, $g_{\!R}^{[\xi]}$ and $h_{\!R}^{[\xi]}$.
The detailed of renormalization for each channel are presented in Appendix A.
Consequently,
the EFT scattering amplitude for the channels $\xi\!=\!$ $^3\!S_1$,
$^3\!P_0$, $^3\!P_1$, and ${^3\!P_2}$  up-to-NLO can be written as
\begin{eqnarray}\label{eq:24}
T^{[\xi]}_{CS}(p)&=&-\frac{2\pi}{\mu}\frac{ C_0^2(\eta_p)W_l(p) }{\frac{(2l+1)2\pi\mathrm{\Delta}_{\!R}^{[\xi]}}{\eta^{[\xi]}g_{\!R}^{[\xi]^2}\!\mu}-\frac{1}{2}(\frac{(2l+1)2\pi }{\eta^{[\xi]}g_{\!R}^{[\xi]^2}\!\mu^2})p^2\!-\! H_l(\eta_p)}
\nonumber\\ &&\times \Big[\!\underbrace{\!\!1_{_{_{_{_{_{_{_{_{_{_{_{_{_{_{_{_{_{_{}}}}}}}}}}}}}}}}}}}\!\!\!\!\!}_{\mathrm{LO}}\!\!+\underbrace{\frac{1}{4}\frac{ (\frac{(2l+1)2\pi h_{\!R}^{[\xi]}}{g_{\!R}^{[\xi]^2}\!\mu^3})}{\frac{(2l+1)2\pi\mathrm{\Delta}_{\!R}^{[\xi]}}{\eta^{[\xi]}g_{\!R}^{[\xi]^2}\!\mu}\!-\!\frac{1}{2}(\frac{(2l+1)2\pi }{\eta^{[\xi]}g_{\!R}^{[\xi]^2}\!\mu^2})p^2\!-\!H_l(\eta_p)}p^4}_{\mathrm{NLO~corection}} \Big]\!.\nonumber\\
\end{eqnarray}
and for the channels $\xi={^3\!D_1}, {^3\!D_2}$, and ${^3\!D_3}$, we have
the LO scattering amplitude as
\begin{eqnarray}\label{eq:25}
T^{[\xi]}_{CS}(p)=-\frac{2\pi}{\mu}\!\frac{ C_0^2(\eta_p)W_l(p) }{\frac{(2l+1)2\pi\mathrm{\Delta}_{\!R}^{[\xi]}}{\eta^{[\xi]}g_{\!R}^{[\xi]^2}\!\mu}\!-\! \frac{1}{2}(\frac{(2l+1)2\pi }{\eta^{[\xi]}g_{\!R}^{[\xi]^2}\!\mu^2})p^2\!-\! \frac{1}{4}(\frac{(2l+1)2\pi h_{\!R}^{[\xi]}}{g_{\!R}^{[\xi]^2}\!\!\mu^3})p^4\!-\! H_l(\eta_p)}.\nonumber\\
\end{eqnarray}
In the other words, according to Eq.~(\ref{eq:12}) the ERE
scattering amplitude corresponding to the EFT scattering amplitudes of Eqs.~(\ref{eq:24}) and (\ref{eq:25}) for $\xi\!=\!{^3\!S_1}, {^3\!P_0}, {^3\!P_1}$, and ${^3\!P_2}$ channels is
\begin{eqnarray}\label{eq:26}
 T^{[\xi]}_{CS}(p)&=&-\frac{2\pi}{\mu}\frac{C_0^2(\eta_p)W_l(p) }{-\frac{1}{a^{[\xi]}}+\frac{1 }{2}r^{[\xi]}p^2- H_l(\eta_p)}  \Big[\underbrace{\,\,\,1_{_{_{_{_{_{_{_{_{_{_{}}}}}}}}}}}}_{\mathrm{LO}}-\underbrace{\frac{1}{4}\frac{s^{[\xi]}}{-\frac{1}{a^{[\xi]}}+\frac{1 }{2}r^{[\xi]}p^2- H_l(\eta_p)}p^4}_{\mathrm{NLO~corection}}\Big],\quad\quad\,\,
\end{eqnarray}
and in $\xi\!=\!{^3\!D_1}, {^3\!D_2}$ and ${^3\!D_3}$ channels is
\begin{eqnarray}\label{eq:27}
T^{[\xi]}_{CS}(p)=-\frac{2\pi}{\mu}\frac{C_0^2(\eta_p)W_l(p)
}{-\frac{1}{a^{[\xi]}}+\frac{1
}{2}r^{[\xi]}p^2+\frac{1}{4}s^{[\xi]}p^4-H_l(\eta_p)},\quad\quad
\end{eqnarray}
Comparing Eqs.~(\ref{eq:24}) and (\ref{eq:25}) with
(\ref{eq:26}) and (\ref{eq:27}) yields
\begin{eqnarray}\label{eq:28}
\mathrm{\Delta}_{\!R}^{[\xi]}&=&-\frac{\mu\eta^{[\xi]}g_{\!R}^{[\xi]^2}}{(2l+1)2\pi a^{[\xi]}},\\
g_{\!R}^{[\xi]^2}&=&-\frac{(2l+1)2\pi}{\mu^2\eta^{[\xi]}r^{[\xi]}},\\
h_{\!R}^{[\xi]}&=&-\frac{\mu^3g_{\!R}^{[\xi]^2} s^{[\xi]}}{(2l+1)2\pi}.
\end{eqnarray}
Although the unknown EFT LECs $g^{[\xi]}$, $\triangle^{[\xi]}$ and $h^{[\xi]}$ are regularization scheme dependent and can not be directly measured but their renormalized EFT LECs $g_{\!R}^{[\xi]}$, $\triangle_{\!R}^{[\xi]}$ and $h_{\!R}^{[\xi]}$  and also sign
of the parameter $\eta^{[\xi]}$ should be initially determined by matching EFT expression of phase shifts
to the available experimental data as we explain in the next section.

In summary, the LO and NLO EFT amplitudes for each
partial wave are constructed as follows: For the D waves
($^3D_1$, $^3D_2$, $^3D_3$), because of containing the momentumindependent,
momentum-squared and momentum-cubed divergences
in the propagators, we should consider all three
parameters a, r and s at LO to renormalize the interacting
D-wave propagators via introducing the renormalized EFT
LECs. For the P waves ($^3P_0$, $^3P_1$, $^3P_2$), since the propagators
contain the momentum-independent and momentumsquared
divergences, we need to consider two parameters
a and r at LO to renormalize the interacting P-wave propagators
via introducing the renormalized EFT LECs and the
shape parameter s is entered at NLO. However, according to
our suggested PC which is represented in the next section, it
can be seen that the second and third terms (effective range
and shape parameter) behave as higher order correction compared
to the first term (scattering length). For the $^3S_1$ wave,
the propagator has only the momentum-independent divergence.
So, considering of the first term (scattering length) is enough for the renormalization. But according to our suggested
PC, the second term (effective range) in this channel
is three orders smaller than the first term. Therefore, for simplifying
themanuscript and matching the formulation of EFT
amplitude for S wave with P waves, we have considered two
parameters a and r at LO same as P channels.

\section {EFT coupling constants determination}
\begin{figure*}
  \begin{center}
\includegraphics[width=6in,height=19.0cm]{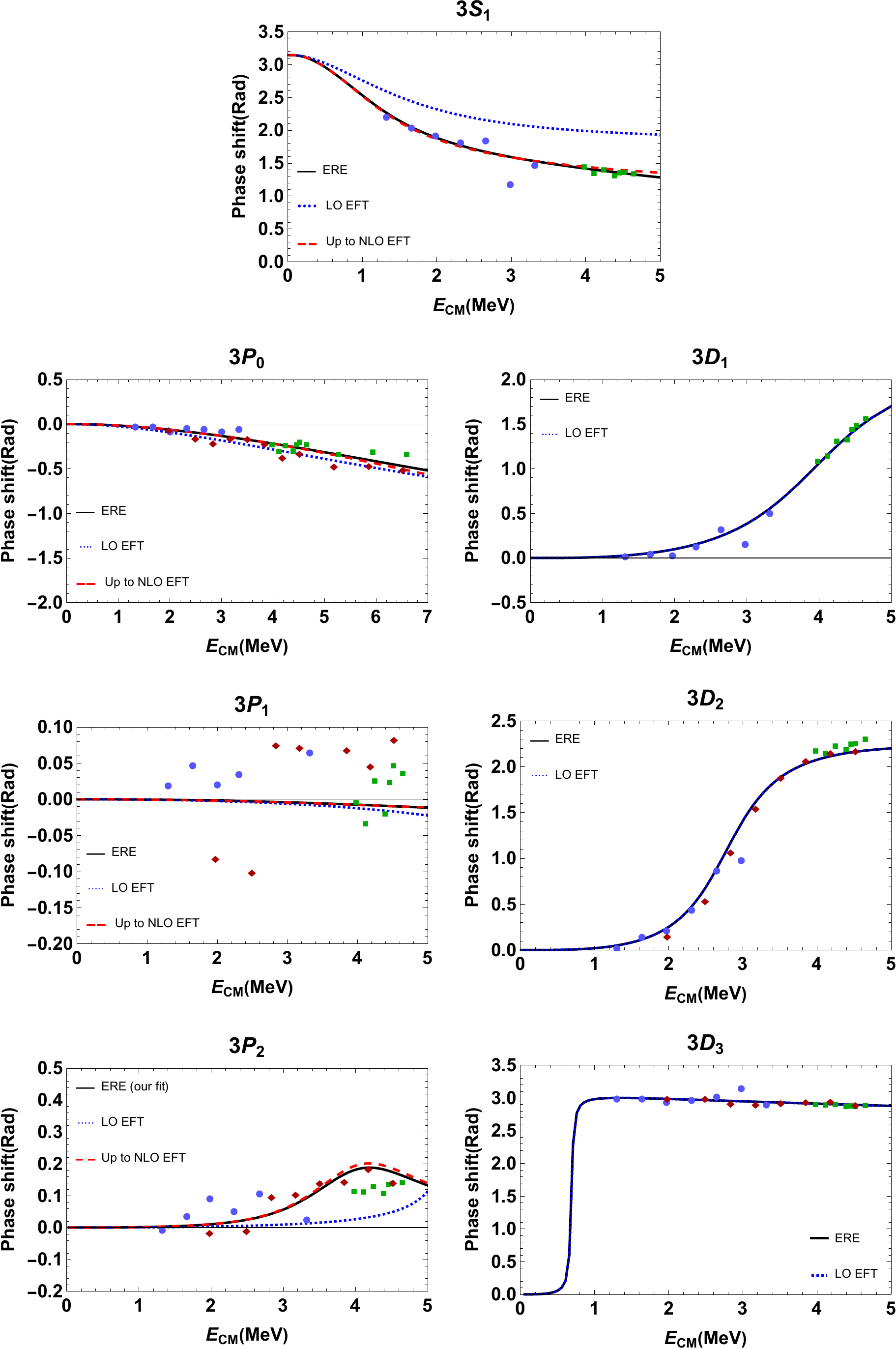}
\end{center}
\caption{\small{Comparison of the ERE and our two-body cluster EFT
fits for the $d-\alpha$ scattering phase shift. The blue-dotted,
red-dashed and black solid lines represent LO EFT, up-to-NLO EFT and
ERE results, respectively. Circles \cite{gruebler1975phase}, squares \cite{A. Jenny}, and
diamonds \cite{schmelzbach1972phase} are the experimental data. According to the
described scheme in the Sec.~\ref{sec:2}, we consider the influences of all
three scattering length, effective range, and shape parameters of the
$D$-wave channels simultaneously, so we have only single LO plot for
the $D$ waves}.}\label{phase}
\end{figure*}

{\label{sec:3}}
As previously explained, in the low-energy $d-\alpha$ scattering the
$S$-, $P$-, and $D$-wave channels ($\xi\! =\!{^3\!S_1}, {^3\!P_0}, {^3\!P_1},
{^3\!P_2}$, $ {^3\!D_1}, {^3\!D_2}$, and $ {^3\!D_3}$) dominantly contribute in
the scattering cross section. Calculating the physical
scattering observables e.g., phase shifts and cross section based on
our EFT expressions, needs to determine the values of the LECs in
the Lagrangian (\ref{eq:17}). This constructed cluster EFT for the
$d-\alpha$ system is reliable at the incident CM energies below $3.3$
MeV. A low-energy phase shift analysis was frequently reported for
the elastic scattering in Refs. \cite {A. Jenny,schmelzbach1972phase,gruebler1975phase}. The existing phase shift
data help us to obtain the values of EFT LECs for all channels.
Taking into consideration Eq.~(\ref{eq:11}), the phase shifts for
each partial waves is obtained from
\begin{eqnarray}\label{eq:30}
\delta^{[\xi]}(p)&=&\textrm{cot}^{-1} \bigg\{-\frac{2\pi}{\mu p}\mathrm{Re}\big[(T^{[\xi]}_{CS}(p))^{-1}\big]\bigg\}.
\end{eqnarray}
Matching Eq. (\ref{eq:30}) with the scattering amplitudes in Eqs. (\ref{eq:12}), (\ref{eq:26}) and (\ref{eq:27}) to the
available low-energy phase shift data \cite {A. Jenny,schmelzbach1972phase,gruebler1975phase} for all possible
channels $\xi$, the values of the effective range
parameters are obtained. The fitted plots of the $d-\alpha$
scattering phase shifts are shown in Fig.~\ref{phase}. Regarding our
suggested scheme, the LO (up to NLO) EFT and ERE results of all
$S$-, $P$-, and $D$-wave phase shifts are plotted against CM energy
by dotted (dashed) and solid lines, respectively. The circles \cite{gruebler1975phase},
squares \cite{A. Jenny} and diamonds \cite {schmelzbach1972phase} indicate the available low-energy
experimental data.
\renewcommand{\arraystretch}{1.28}
\begin{table*}
\caption{\small{The determined effective range parameters. The
parameters were obtained from matching the LO (up-to-NLO) EFT and ERE relations
to the available low-energy
experimental data in Refs.\cite {A. Jenny,schmelzbach1972phase,gruebler1975phase} for each channel $\xi=$$^3\!S_1$,
$^3\!P_0$, $^3\!P_1$, $^3\!P_2$, $^3\!D_1$, $^3\!D_2$, and $^3\!D_3$ as shown in Fig.~\ref{phase}. The last
column shows the deviations of fits from phase shifts experimental
data}.}{\label{table:1}}
\begin{center}
    \begin{tabular}{cccccc}
  \hline
$\xi$& Method
&~~$\small{a^{[\xi]}[{\textrm{MeV}^{-2l-1}}]}$&~~~$\small{r^{[\xi]}[{\textrm{MeV}^{2l-1}}]}$
&~$\small{s^{[\xi]}[{\textrm{MeV}^{2l-3} }]}$
&${\chi^{[\xi]}}^2$\\
 \hline
& LO EFT&$\small{-2.060\times 10^{-2}}$&$\,\,\,\,\,\,\,\,3.533 \times 10^{-3}$&$-$&$2.6187$\\
$^3\!S_1$&NLO EFT&$-2.780\times 10^{-2}$&$\,\,\,\,\,\,\,\,3.830 \times 10^{-3}$&$-8.345 \times 10^{-7}$&$0.0775$\\
&ERE&$-2.757\times 10^{-2}$&$\,\,\,\,\,\,\,\,3.137 \times 10^{-3}$&$-7.688 \times 10^{-7}$&$0.0283$\\
 \hline
&LO EFT&$-8.029\times 10^{-7}$&$ \,\,\,\,\, 2.163 \times 10^{2}$&$-$&$0.0069$\\
$^3\!P_0$&NLO EFT&$-7.824\times 10^{-7}$&$\,\,\,\,\,1.356 \times 10^{2}$&$\,\,\,\,\,1.950 \times 10^{-3}$&$0.0014$\\
&ERE&$-4.364\times 10^{-7}$&$\,\,\,\,\,1.496 \times 10^{2}$&$\,\,\,\,\,1.634 \times 10^{-3}$&$0.0001$\\
%$$&$$\textrm{PC estimation}$$&$6.561\times 10^{6}$&$9.000\times 10$&$1.234\times 10^{-3}$&$$\\
 \hline
&LO EFT&$-2.161\times 10^{-8}$&$-6.166 \times 10^{3}$&$-$&$0.0433$\\
$^3\!P_1$&NLO EFT&$-1.004\times 10^{-8}$&$-7.494\times 10^{3} $&$\!\!\!\!\!\!\!\!\!\!\!\!\!\!0.474 $&$0.0024$\\
&ERE&$-1.012\times 10^{-8}$&$-8.494\times 10^{3} $&$\!\!\!\!\!\!\!\!\!\!\!\!\!\!0.452 $&$0.0021$\\
%$$&$$\textrm{Value}$$&$5.904\times 10^{7}$&$8.100\times 10^{2}$&$1.000\times 10^{-1}$&$-$\\
 \hline
&LO EFT&$ \,\,\,\,\,1.297\times 10^{-8}$&$\,\,\,\,\, 1.124 \times 10^{5}$&$-$&$1.4657$\\
$^3\!P_2$&NLO EFT&$\,\,\,\,\, 2.014\times 10^{-8}$&$\,\,\,\,\, 1.874\times 10^5 $&$\!\!\!\!\!\!\!\!\!\!\!\!\!\!\!\!\!\!-1.851$&$1.8406$\\
&ERE&$\,\,\,\,\, 2.037\times 10^{-8}$&$\,\,\,\,\, 1.864\times 10^5 $&$\!\!\!\!\!\!\!\!\!\!\!\!\!\!\!\!\!\!-1.865$&$1.5520$\\
%$$&$$\textrm{Value}$$&$5.904\times 10^{7}$&$\quad 6.561\times 10^{4}$&$9.000\times 10^{-1}$&$-$\\
 \hline
$^3\!D_1$&LO EFT/ERE&$\,\,-1.375\times 10^{-10}$&$\,\,\,\,\, 1.012\times 10^{6}$&$\!\!-1.905 \times 10^{3}$&$0.5597$\\
%$$&$$\textrm{Value}$$&$5.904\times 10^{9}$&$7.290\times 10^{5}$&$8.100\times 10^2$&$-$\\
 \hline
$^3\!D_2$&LO EFT/ERE&$\,\,-1.716\times 10^{-10}$&$-7.086 \times 10^ 5 $&$\!\!\!\!\!\!\!\!\!\!\!\!\!\!\!\!\!\!-10.958 $&$0.0033$\\
%$$&$$\textrm{Value}$$&$5.904\times 10^{9}$&$8.100\times 10^{4}$&$1.000\times 10$&$$\\
 \hline
$^3\!D_3$&LO EFT/ERE&$-4.500\times 10^{-8}$&$-1.554\times 10^{6}$&$1.303 \times 10^{3}$&$0.0028$\\
%$$&$$\textrm{Value}$$&$7.290\times 10^{7}$&$7.290\times 10^{5}$&$8.100\times 10^2$&$-$\\
 \hline
    \end{tabular}
\end{center}
\end{table*}
The determined effective range parameters of channel $\xi$ has been reported in Table \ref{table:1}. The quality of
description of available results $f^{ave}$ on the basis of the
certain expression $f$ can be estimated by the $\chi^2$ method which
is written as~\cite {MMA-dt-dt}
\begin{eqnarray}{\label{eq:31}}
\chi^2=\frac{1}{N}\sum^N_{i=1}\Big[\frac{f_i-f^{ave}}{f^{ave}}  \Big]^2,
\end{eqnarray}
where $N$ is the number of measurements. Taking into consideration
$f$ as $\delta^{[\xi]}$ introduced in Eq. (\ref{eq:30}), the deviations of fits from used phase shift data for
$\xi$ channel are obtained as shown in the last column of
Table~\ref{table:1}.

The phase shift analysis in Fig.~\ref{phase} leads to the
effective-range parameters presented in Table~\ref{table:1}. Based
on determined values from ERE fits, we propose a power-counting (PC) in
which the effective-range parameters of $\xi$ channel are scaled as
presented in Table~\ref{table:2}. So, we can conclude that the main
contribution of the scattering amplitude in all channels
$^3\!S_1$, $^3\!P_0$, $^3\!P_1$, $^3\!P_2$, $^3\!D_1$ and $^3\!D_2$ come clearly
from their scattering lengths, and the influences of both their
effective ranges and shape parameters are small and can be considered as higher-order
corrections. In this analysis, the effective-range and
shape-parameter terms are suppressed by $(Q/\mathrm{\Lambda})^n$ and
$(Q/\mathrm{\Lambda})^m$ as compared to the leading term of the $^3\!S_1$,
$^3\!P_0$, $^3\!P_1$, $^3\!P_2$, $^3\!D_1$ and $^3\!D_2$ with
$n=3,~3,~2,~1,~2,~3$ and $m=5,~7,~5,~4,~3,~5$, respectively.

For the $^3\!D_3$ partial wave, it seems that the contribution of both scattering length and shape parameters in comparison with the effective  range term are one order down. However, no missing any physical effect,
we would
consider $-\frac{1}{a}+\frac{1}{2}r p^2+\frac{1}{4}s p^4\sim
Q^2\mathrm{\Lambda}^3$ in the leading order.
Furthermore, in the case corresponds to the large value of $\eta_p$,
the term $H_l(\eta_p)$ is significantly different from the usual
unitary term $ip$. Therefore, in this case, the unitary term leads
to $H(\eta_p)\!\sim\! p^2/12k_C^2$ \cite{Higa-Hammer}. For the $S$-wave
channels, $H_0(\eta_p)$ is comparable in magnitude to the
effective-range term and can be automatically captured by taking
$3k_C\sim \mathrm{\Lambda}$. Alternatively, one can enhance by a factor of
$\mathrm{\Lambda}/Q$ the size of the $^3S_1$ effective range. In the $P$ waves, we have
$H_1(\eta_p)\!\sim \!Q^3(1\!+\!Q/\mathrm{\Lambda})$ and the term
including $H(\eta_p)$ can be also managed by redefining the
effective range and shape parameter \cite{MMA-dt-dt}. Scaling
$\frac{1}{24}k_C^3\!\sim \!Q^3$, $\frac{5}{24}k_C\!\sim\! Q$ and $6k_C\!\sim\!
\mathrm{\Lambda}$, the function $H_l(\eta_p)$ can be estimated for the $l\!=\!2$
partial waves as $H_2(\eta_p)\sim
Q^5(1+1+Q/\mathrm{\Lambda})$. So, for the $D$ waves, the functions of $p^2$
and $p^4$ can be captured by the effective range and shape
parameter, respectively, and the term regarding the $p^6$ would be
negligible in the current theory.
\renewcommand{\arraystretch}{1.28}
\begin{table}
\caption{\small{The suggested power-counting for the effective range
parameters. $Q$ and $\mathrm{\Lambda}$ denote the low- and
high-momentum scales as introduced in the text}.}{\label{table:2}}
\begin{center}
    \begin{tabular}{cccc}
 \hline
$[\xi]$&$ 1/a^{[\xi]}$&$ r^{[\xi]}/2$ &$ s^{[\xi]}/4$ \\
\hline
%$$&${\Lambda}$&$-$&$-$\\
%&$\textrm{N}^3\textrm{LO}$&${\Lambda}$&${Q}/{\Lambda^2}$&$-$\\
$^3\!S_1$&${\mathrm{\Lambda}}$&${Q}/{\mathrm{\Lambda}^2}$&${Q}/{\mathrm{\Lambda}^4}$\\
%$$&$$\textrm{Value}$$&$90$&$1.234\times 10^{-3}$&$1.524\times 10^{-7}$\\
\hline
%$$&${\Lambda^4}/Q$&$-$&$-$\\
%$^3P_0$&$\textrm{N}^3\textrm{LO}$&${\Lambda^4}/Q$&$\Lambda$&$-$\\
$^3\!P_0$&${\mathrm{\Lambda}^4}/Q$&$\mathrm{\Lambda}$&${Q^2}/{\mathrm{\Lambda}^3}$\\
%$$&$$\textrm{Value}$$&$6.561\times 10^{6}$&$9.000\times 10$&$1.234\times 10^{-3}$\\

%$$&$\Lambda^{5}/Q^2$&$-$&$-$\\
%$^3P_1$&$\textrm{N}^3\textrm{LO}$&$\Lambda^{5}/Q^2$&${\Lambda^2}/{Q}$&$-$\\
$^3\!P_1$&$\mathrm{\Lambda}^{5}/Q^2$&${\mathrm{\Lambda}^3}/{Q^2}$&${1}/{Q}$\\
%$$&$$\textrm{Value}$$&$5.904\times 10^{7}$&$8.100\times 10^{2}$&$1.000\times 10^{-1}$\\

%$$&$\Lambda^{5}/Q^2$&$-$&$-$\\
%$[^3P_2]$&$\textrm{N}^3\textrm{LO}$&$\Lambda^{5}/Q^2$&${\Lambda^4}/{Q^3}$&$-$\\
$^3\!P_2$&$\mathrm{\Lambda}^{5}/Q^2$&${\mathrm{\Lambda}^4}/{Q^3}$&${\mathrm{\Lambda}}/{Q^2}$\\
%$$&$$\textrm{Value}$$&$5.904\times 10^{7}$&$6.561\times 10^{4}$&$9.000\times 10^{-1}$\\
\hline
%$$&$\Lambda^{5}$&$-$&$-$\\
%$[^3D_1]$&$\textrm{N}^2\textrm{LO}$&$\Lambda^{5}$&$\Lambda^3$&$-$\\
$^3\!D_1$&$\mathrm{\Lambda}^{5}$&$\mathrm{\Lambda}^3$&${\mathrm{\Lambda}^2}/{Q}$\\
%$$&$$\textrm{Value}$$&$5.904\times 10^{9}$&$7.290\times 10^{5}$&$8.100\times 10^2$\\

%$$&$\Lambda^{5}$&$-$&$-$\\
%$[^3D_2]$&$\textrm{N}^3\textrm{LO}$&$\Lambda^{5}$&$Q\Lambda^2$&$-$\\
$^3\!D_2$&$\mathrm{\Lambda}^{5}$&$Q\mathrm{\Lambda}^2$&$Q$\\
%$$&$$\textrm{Value}$$&$5.904\times 10^{9}$&$8.100\times 10^{4}$&$1.000\times 10$\\

%$$&$Q^2\Lambda^3$&$\Lambda^{3}$&$-$\\
$^3\!D_3$&$Q^3\mathrm{\Lambda}^2$&$\mathrm{\Lambda}^{3}$&${\mathrm{\Lambda}^2}/{Q}$\\
%$$&$$\textrm{Value}$$&$7.290\times 10^{7}$&$7.290\times 10^{5}$&$8.100\times 10^2$\\
\hline
    \end{tabular}
\end{center}
\end{table}

Taking into consideration the LO and NLO values of effective range
parameters corresponding to the scheme used in Table~\ref{table:1},
the LO and NLO values of EFT LECs for channel $\xi$ are determined
as indicated in the first and second rows of Table \ref{table:3}.
Based on the suggested PC in Table~\ref{table:2}, the
estimation of the LECs for each channel are presented as "PC
estimation" in Table \ref{table:3}. The orders of obtained EFT LECs
are meaningfully consistent with the predictions of the suggested PC.
\section{Differential Cross Section}\label{sec:4}
In this section, we present the obtained results of the $d-\alpha$
differential cross section in the two-body cluster EFT approach. The
differential cross section for the $d-\alpha$ elastic scattering
with the contributions of the Coulomb and the strong interactions is
given by
\begin{equation}\label{eq:32}
\frac{d\sigma}{d\Omega}=\Big(\frac{\mu}{2\pi}\Big)^2 |T_C+T_{CS}|^2.
\end{equation}
\renewcommand{\arraystretch}{1.28}
\begin{table*}
\caption{The obtained EFT coupling constants for all $l=0,~1,~2$
channels using the determined effective-range parameters in
Table~\ref{table:1}. The LO and NLO results are consistent with the
suggested scheme as introduced in the Sec.~\ref{sec:2}. The last row in each
channel states our suggested PC estimation based on Table
\ref{table:2}.}{\label{table:3}}
\begin{center}
    \begin{tabular}{ccccc}
\hline

   $\xi$&Order& $\quad\small{\mathrm{\Delta}_{\!R}^{[\xi]}}[\textrm{MeV}]$ &\qquad\quad\quad\quad $\small{g_{\!R}^{[\xi]}}[\textrm{MeV}^{-(2l+1)/2}] $&\quad $\small{h_{\!R}^{[\xi]}}[\textrm{MeV}^{-1}]$ \\
\hline
 $$&{LO}&$ \!\!-10.944 $&\qquad\quad\quad\quad$3.360\times 10^{-2}$&\quad$-$\\
$^3\!S_1$&NLO&$-7.467$&\qquad\quad\quad\quad$3.231 \times 10^{-3}$&\quad$0.272$\\
$$&$\textrm{PC estimation}$&\!\!\!\!\!\!\!\!\!\!\!\!\!\!$\frac{\mathrm{\Lambda}^3 }{2\mu Q}\!= \!16.175 $&\qquad\!\!\!\!\!\!\!\!\!$(\frac{\pi \mathrm{\Lambda}^2}{\mu^2 Q})^{\frac{1}{2}}\!=3.003\times 10^{-2}$&\!\!\!\!\!\!\!\!\!$\frac{ 2\mu}{\mathrm{\Lambda}^2}\!=0.309$\\
\hline
 $$&{LO}&$-4.577$&\qquad\quad\quad\quad$2.352\times 10^{-4}$&\qquad\!\!\!\!\!$-$\\
$^3\!P_0$&NLO&$ \!\!-13.091$&\qquad\quad\quad\quad$2.977 \times 10 ^{-4}$&$\qquad\quad\,-1.801 \times 10^{-2}$\\
$$&$$\textrm{PC estimation}$$&\!\!\!\!\!\!\!\!\!\!\!\!\!\!$\frac{\mathrm{\Lambda}^3 }{2\mu Q}\!=  \!16.175 $&\qquad\!\!\!\!\!\!\!\!$(\frac{3\pi }{\mu^2 \mathrm{\Lambda}})^{\frac{1}{2}}\!=2.584\times 10 ^{-4}$&\qquad\!\!\!\!\!\!\!\!$\frac{ 2\mu Q^2}{\mathrm{\Lambda}^4}\!=1.236\times 10^{-2}$\\
\hline
$$&{LO}&$\,\,\,\,5.992$&\qquad\quad\quad\quad$4.416 \times 10^{-5}$&\,\,\,\,$-$\\
$^3\!P_1$&NLO&$ \, \,10.607 $&\qquad\quad\quad\quad$4.006 \times 10 ^{-5}$&$\qquad\quad\,\,\,\,7.930 \times 10^{-2}$\\
$$&$$\textrm{PC estimation}$$&\!\!\!\!\!\!\!\!\!\!\!$\frac{\mathrm{\Lambda}^2 }{2\mu }\!= \!\,\,\,3.235$&\qquad\!\!\!\!\!\!\!\!\!\!$(\frac{3\pi Q^2}{\mu^2 \mathrm{\Lambda}^3})^{\frac{1}{2}}\!=5.169 \times 10^{-5}$&\qquad\!\!\!\!\!\!\!$\frac{ 2\mu Q}{\mathrm{\Lambda}^3}\!=6.182\times 10^{-2}$\\
\hline
$$&{LO}&$\,\,\,\,\,5.474$&\qquad\quad\quad\quad$3.269 \times 10^{-5}$&\,\,$-$\\
$^3\!P_2$&NLO&\,\,\, 2.114&\qquad\quad\quad\quad$2.532\times 10^{-5}$&$\,\,0.123$\\
$$$$&$$\textrm{PC estimation}$$&\!\!\!\!\!\!\!\!\!\!\!\!$\frac{Q\mathrm{\Lambda} }{2\mu }\!= \!\,\,\,0.647$&\qquad\!\!\!\!\!\!\!\!\!\!$(\frac{3\pi Q^3}{\mu^2 \mathrm{\Lambda}^4})^{\frac{1}{2}}\!=2.312\times 10^{-5}$&\!\!\!\!\!\!\!\!\!\!\!\!\!\!\!$\frac{ 2\mu Q}{\mathrm{\Lambda}^3}\!=0.062$\\
\hline
$^3\!D_1$&LO&$-5.730$&\qquad\quad\quad\quad$4.448\times 10^{-6}$&$\,\,\,\,\!\!\,0.235$\\
$$&$$\textrm{PC estimation}$$&\!\!\!\!\!\!\!\!\!\!\!\!\!$\frac{\mathrm{\Lambda}^2 }{2\mu }\!= \,\,3.235$&\qquad\!\!\!\!\!\!\!\!\!$(\frac{5\pi }{\mu^2 \mathrm{\Lambda}^3})^{\frac{1}{2}}\!=3.707\times 10^{-6}$&\!\!\!\!\!\!\!\!\!\!\!\!\!$\frac{ 2\mu }{\mathrm{\Lambda}Q}\!=1.545$\\
\hline
$^3\!D_2$&LO&$\,\,\,\,\, 6.568$&\qquad\quad\quad\quad\,$5.318\times 10^{-6}$&\!\!\!\!\!\,\,\,\,\!\!\,$-0.017 $\\
$$&$$\textrm{PC estimation }$$&\!\!\!\!\!\!\!\!\!\!\!\!\!\!$\frac{\mathrm{\Lambda}^3 }{2\mu Q}\!= \!16.175 $&\qquad\!\!\!\!\!\!\!\!\!\!\!\!$(\frac{5\pi }{\mu^2 \mathrm{\Lambda}^2Q})^{\frac{1}{2}}\!=\!8.291\times 10^{-6}$&\!\!\!\!\!\!\!\!\!\!\!\!\!$\frac{ 2\mu }{\mathrm{\Lambda}^2}\!=0.309 $\\
\hline
$^3\!D_3$&LO&$\quad\quad\quad\quad1.578 \times 10^{-2}$&\qquad\quad\quad\quad$1.888\times 10^{-6}$&$1.450$\\
$$&$$\textrm{PC estimation}$$&\,\,\,\,\,\quad\quad\,\,\!\!\!\!\!\!\!\!\!\!\!\!\!$\frac{Q^3 }{2\mu \mathrm{\Lambda}}\!=\!\!\,\,\,2.588 \times 10^{-2}$&\qquad\!\!\!\!\!\!\!\!\!$(\frac{5\pi }{\mu^2 \mathrm{\Lambda}^3})^{\frac{1}{2}}\!=\!3.707\times 10^{-6}$&\!\!\!\!\!\!\!\!\!\!\!\!\!\!$\frac{ 2\mu }{\mathrm{\Lambda}Q}\!=\!1.545$\\
\hline
    \end{tabular}
\end{center}
\end{table*}
Taking into account the determined values of EFT LECs presented in
Table~\ref{table:3}, we can compute the differential cross section
at different CM energies and scattering angles. In order to
calculate the differential cross section for the low-energy
$d-\alpha$ elastic scattering, some important issues should be
clarified. At the low energies, the cross section gets the dominant
contribution from the leading term of the scattering amplitude in
the $^3$$\!S_1$ partial wave. {
Thus, regarding the phase shift
analysis for all $S$-, $P$- and $D$-wave channels in
Tables~\ref{table:1} and \ref{table:2}, the leading $d-\alpha$
scattering cross section constructed by the relation corresponding
to the scattering length of $^3$$\!S_{1}$ channel.}

Based on our analysis in the previous section, the biggest
corrections on the LO cross section comes from the
effective range of $^3$$\!S_{1}$ and also the scattering length and
effective range of $^3$$\!D_{3}$ partial wave corresponding to the
first four terms of Lagrangian (\ref{eq:17}). These corrections are
two orders down with respect to the effect of the $^3$$\!S_{1}$
scattering length. Remained effective range parameters could be
neglected as $\textrm{N}^3\textrm{LO}$ and higher-order
contributions in the current calculation.

\begin{figure*}
 \begin{center}
 \includegraphics[width=6in,height=7.4in]{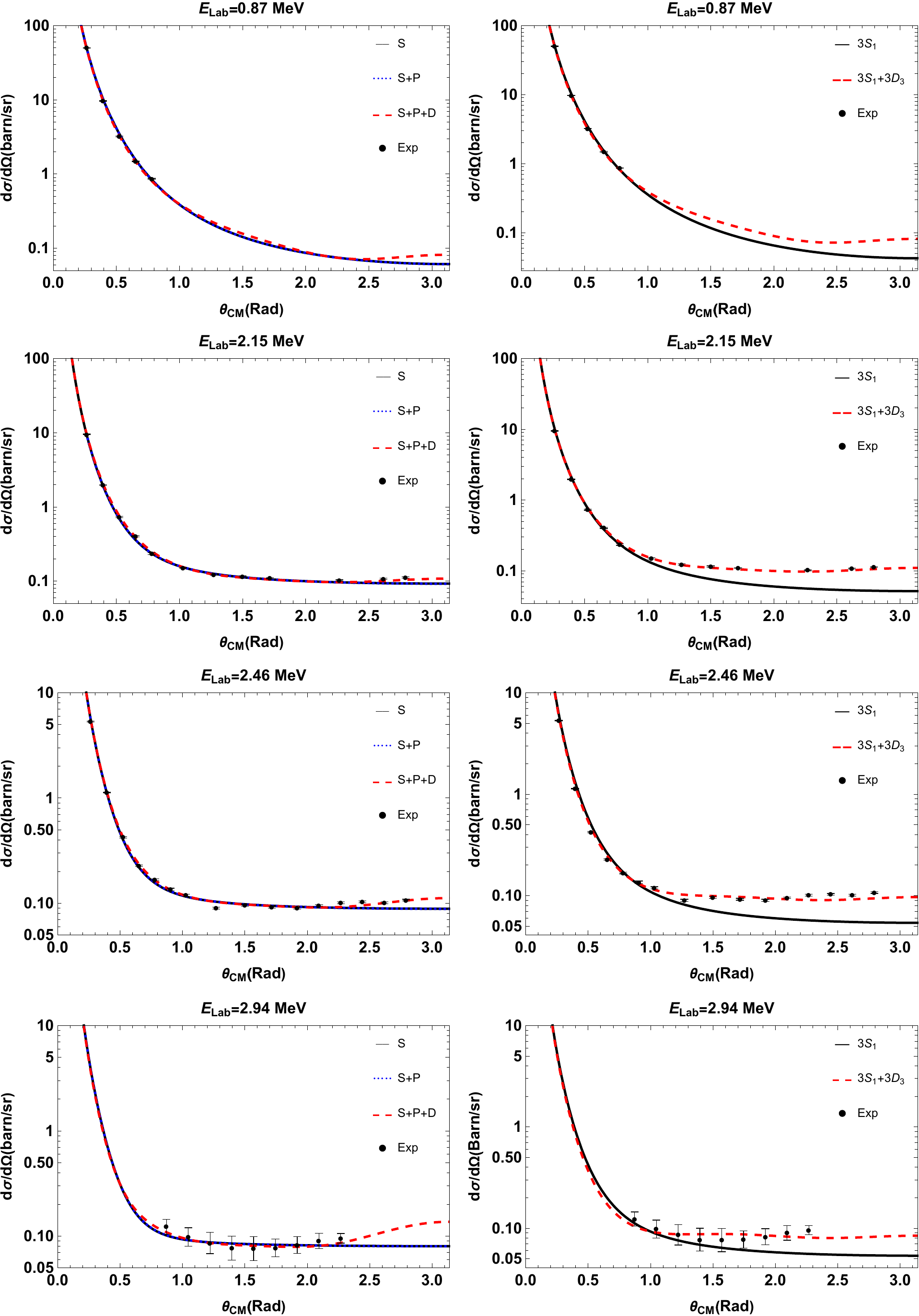}
 \end{center}
%\begin{flushleft}
\caption{\small{Differential cross sections for the low-energy $d-\alpha$
elastic scattering against the CM angle. Our EFT results are
plotted with the laboratory energies $E_{Lab}=0.87$, 2.15, 2.46, and
2.94 MeV. The left column shows the the calculated EFT cross section
with the contribution of the $S$ (black-solid), $S$+$P$
(blue-dotted) and $S$+$P$+$D$ (red-dashed). The right column
indicate our plots for the differential cross section using the
leading terms of $^3\!S_1$ only (black-solid), and the leading
scattering terms in $^3\!S_1$ including the effects of $^3\!D_3$ channel
(red-dashed). The dots are the experimental data from Refs.} \cite {ohlsen1964deuteron,blair1949scattering}.
}\label{fig:4}
%\end{flushleft}
\end{figure*}

Our results for the differential cross section versus the CM
scattering angle for the $d-\alpha$ scattering are shown in
Fig.~\ref{fig:4} for the laboratory energies $E_{Lab}=0.87$, 2.15,
2.46, and 2.94 MeV. The contribution of $S$-, $P$- and $D$- waves in the
differential cross section are shown in the first column of
Fig.~\ref{fig:4}. And also, the results of the cross section with
the $^3$$\!S_{1}$ ($^3$$\!S_{1}$ and $^3$$\!D_{3}$) partial wave(s) are
depicted by the dashed (solid) line in the second column of
Fig.~\ref{fig:4}. The symbols in Fig.~\ref{fig:4} indicate the
reported experimental data from Refs.
\cite{ohlsen1964deuteron,blair1949scattering}.

\begin{figure*}
  \begin{center}
 \includegraphics[width=6 in,height=7.4in]{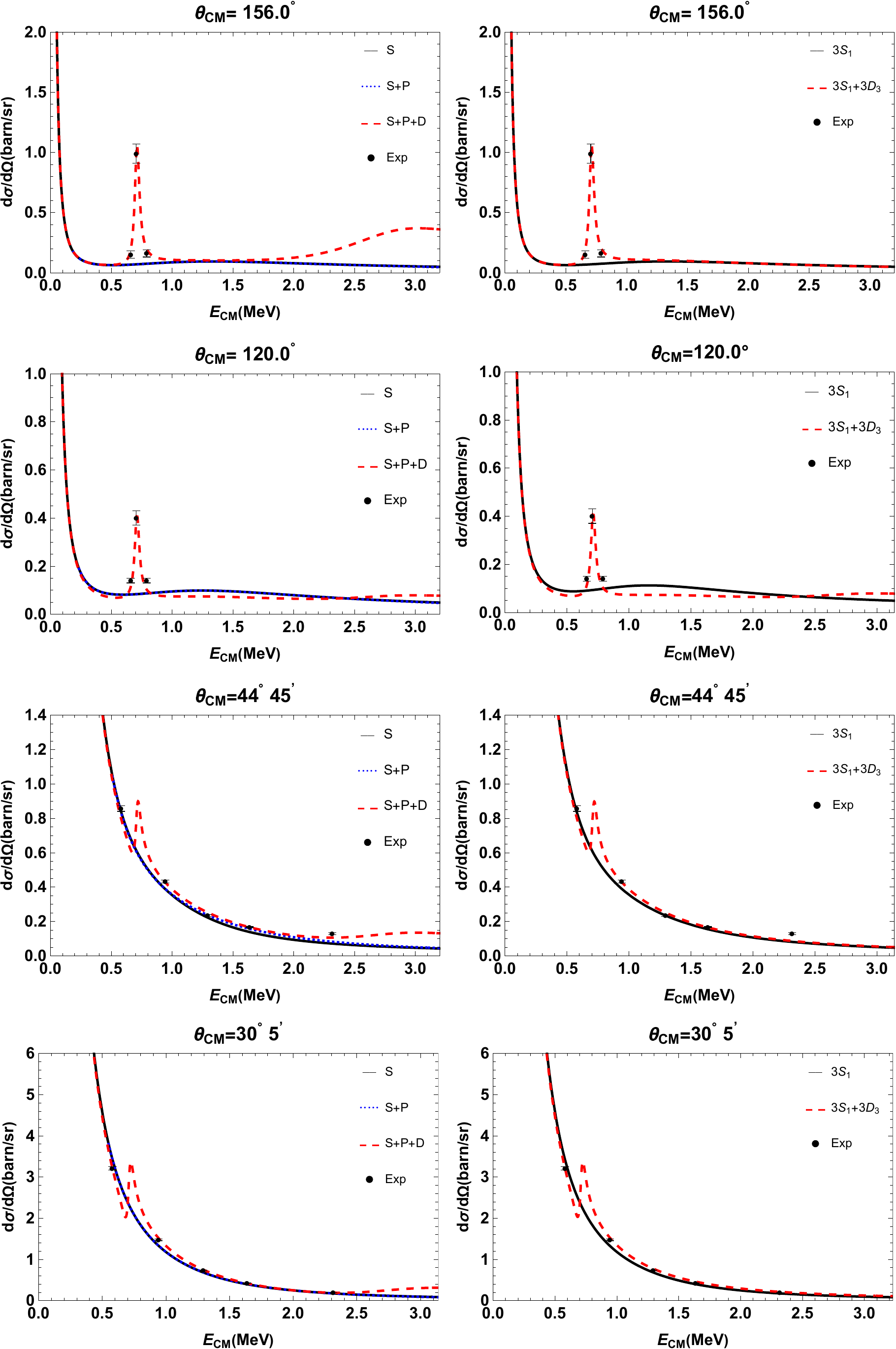}
  \end{center}
 \caption{\small{Differential cross sections for the low-energy $d-\alpha$
elastic scattering against the CM energy. Our EFT results are
plotted with scattering angle $\theta_{CM}\!=\!156^\circ$, $120^\circ$,
$44^\circ {45}'$ and  $30^\circ 5'$. All notations are as in Fig.~\ref{fig:4}}}\label{fig:5}
\end{figure*}

We have also plotted the differential cross sections of the
$d-\alpha$ elastic scattering against CM energy with
scattering angle $\theta_{CM}\!=\!156^\circ$, $120^\circ$,
$44^\circ {45}'$ and  $30^\circ 5'$ in
Fig.~\ref{fig:5}. Our EFT results using the $^3$$\!S_{1}$ ($^3$$\!S_{1}$
and $^3$$\!D_{3}$) channel(s) are depicted by the dashed (solid) line,
and the circles in Fig.~\ref{fig:5} indicate the experimental data
in
Ref.~\cite{ohlsen1964deuteron,blair1949scattering}.
Fig.~\ref{fig:5} shows that in our EFT formalism the peak manner of
the differential cross section around $E_{CM}\sim 0.706 $ MeV can be
reproduced only by including the $^3$$\!D_{3}$ scattering amplitude
with the influences regarding its scattering length and effective
range. It seems that the contributions of the $^3$$\!D_{3}$ would be
more important and it must be included in our EFT calculations to
reproduce reliably the low-energy experimental data.

Our EFT results in Figs.~\ref{fig:4} and \ref{fig:5} indicate that
the $^3$$\!S_{1}$ and $^3$$\!D_{3}$ scattering amplitudes could
reproduce the low-energy experimental data and other partial waves
have no significant effect at the current low-energy regime as we
expected from the suggested PC.

\section{Conclusion}\label{sec:5}

In this paper, we have studied the low-energy $d-\alpha$ elastic
scattering using two-body cluster EFT approach. Our constructed
cluster EFT treats the deuteron and alpha nucleus as the point-like
nuclear clusters, so we have concentrated on the energy region
$E_{CM}\lesssim 3.3$ MeV. At the present energy region, the Coulomb force
has been considered as a non-perturbative treatment. Here, we have
studied all possible $S$-, $P$- and $D$-wave channels. We have
introduced a scheme in which the LO contributions of phase shift in
each partial wave of $l=0,1$ channels has been constructed from its
scattering length and effective range and its shape parameter
influence has been included at the NLO order. Also, the
additional 2nd-order kinetic term with constant $h^{[\xi]}$ is
needed to renormalize the interacting $D$-wave propagator which
contains up to quintic divergences.

Using the available low-energy phase shift data, we obtained the
values of the effective range parameters $S$, $P$ and $D$ waves. The
EFT LECs for $l=0,1,2$ partial waves evaluated in terms of effective
range parameters. Our ERE fitted curves and the cluster EFT
calculations for the $S$-, $P$- and $D$-wave phase shifts have good
consistency with the available results and a converging pattern from
LO to NLO. We have plotted the differential cross sections against
the CM scattering angle and also the CM energy. The comparison
our obtained two-body cluster EFT results to the experimental data
indicates good consistency.

Our obtained EFT results indicate that the cross section of the
$d-\alpha$ scattering got the dominant contributions using the
scattering amplitude of $^3$$\!S_1$ partial wave containing the
dimeron propagator without kinetic energy terms. It regards the
$^3$$\!S_1$ scattering-length effect as we expected from our PC
analysis. We have also showed that the resonance behavior of the
$d-\alpha$ cross section can be reproduced only by including the
contribution of the $^3$$\!D_3$ scattering amplitude. It is consistent
to our PC estimation in which the largest corrections on the leading
$d-\alpha$ scattering cross section are constructed by the strong
interacting contributions corresponding to the $^3$$\!S_1$ effective
range and also $^3$$\!D_3$ scattering length and effective range. It
should be mentioned that other strong interacting terms can be omitted
because of small contributions of orders $\textrm{N}^3\textrm{LO}$
and higher in the total low-energy cross section.

The discrepancy of our results for the cross section above
$E_{CM}>3.3$ MeV can be handled by introducing the three-body cluster
EFT in which neutron, proton and alpha particle are the degrees of
freedom. In the present EFT calculation based on considering the
deuteron as a point-like particle, the EFT results for $E_{CM}>3.3$
MeV are questionable and we should switch to the three-body cluster
formalism for the higher energies.

It would be interesting to use our results for studying of the $d+\alpha$
$\rightarrow^6$$\textrm{Li}+\gamma$ astrophysical radiative capture
based on halo/cluster EFT calculation in the future. The $d-\alpha$
scattering and radiative capture can also be studied by the
three-body EFT formalism for the higher-energy region.

\section*{Acknowledgement}
The authors acknowledge the Iran National Science Foundation (INSF) for financial support.
%%%%%%%%%%%%%%%%%%%%%%%%%%%%%%%%%%%%%%%%%%%%%%%
%

%\clearpage

\appendix 

\section{Derivation of the elastic scattering amplitudes}
In this section, we present the detailed derivation of the $d-\alpha$ elastic scattering amplitudes for all possible partial waves, $l=0,~1,~2$.
\subsection*{$S-$wave channel}{\label{subsec:4}}
According to the Lagrangian~(\ref{eq:17}), the strong interaction in the $\xi={^3\!S_1}$ channel of the $d-\alpha$ system can be described using the up-to-NLO Lagrangian
\begin{eqnarray}{\label{eq:a1}}
\mathcal{L}^{[\xi]}&=&\phi^{\dagger}(i\partial _0+\frac{\nabla ^2}{2m_\alpha})\phi+d_i^{\dagger}(i\partial _0+\frac{\nabla ^2}{2m_d})d_i
\nonumber\\
&&  +\,\eta^{[\xi]}\bar{t}_i^{\,\dagger}\Big{[}i\partial _0\!+\!\frac{\nabla ^2}{2m_t}\!-\!\mathrm{\Delta}^{[\xi]}\Big{]}\bar{t}_i +h^{[\xi]}\bar{t}_i^{\,\dagger}\Big{[}i\partial _0+\frac{\nabla ^2}{2m_t}\Big{]}^2\bar{t}_i\nonumber\\
&& +\,g^{[\xi]}\Big[\bar{t}^{\,\dagger}_i(\phi\, d_i)\!+\!h.c.\Big],
\end{eqnarray}

where $\bar{t}_i$  is the  vector auxiliary field of the ${^3\!S_1}$ dimeron. According to the Feynman diagram of Fig.~\ref{fig:2}, the up-to-NLO EFT scattering amplitude in the $^3\!S_1$ channel can be written as
\begin{eqnarray}{\label{eq:a2}}
-iT_{CS}^{[\xi]}e^{2i\sigma_0}&=&(-ig^{[\xi]})^2 \, \chi_{p'}^{*(-)}(\mathbf{0})\, \varepsilon_j^{d*}\,\varepsilon_j^{\bar{t}}\,iD^{[\xi]}(E,\textbf{0})\varepsilon_i^{\bar{t}*}\, \varepsilon_i^d\,\chi_p^{(+)}(\mathbf{0})\nonumber\\&=&-ig^{[^3S_1]^2}D^{[n,^3S_1]}(E,\textbf{0})\, \varepsilon_j^{d*}\,\varepsilon_j^{t} \,\varepsilon_i^{t*}\, \varepsilon_i^d \, \chi_{p'}^{*(-)}(\mathbf{0})\chi_p^{(+)}(\mathbf{0})\nonumber\\
%&=-ig^{[^3S_1]^2}D^{[n,^3S_1]}(E,\textbf{0})\chi_{p'}^{*(-)}(\mathbf{0})\chi_p^{(+)}(\mathbf{0})\nonumber\\
&=&-ig^{[\xi]^2}\!D^{[\xi]}(E,\textbf{0})W_{0}(\eta_p)C_{0}^2(\eta_p)e^{2i\sigma_0},
\end{eqnarray}
where $\varepsilon_i^{d}$ and $\varepsilon_i^{\bar{t}}$ are polarization vectors of the deuteron and dimeron auxiliary fields respectively, which satisfy the relations
\begin{eqnarray}{\label{eq:a3}}
\varepsilon_j^{\bar{t}*} \,\varepsilon_i^{\bar{t}}=
\delta_{ij},\quad\quad\varepsilon_j^{d*} \,\varepsilon_i^{d}=\frac{1}{3}\delta_{ij}.
\end{eqnarray}
In the last equality of Eq.~(\ref{eq:a2}) we use
\begin{eqnarray}{\label{eq:a4}}
\chi_{p'}^{*(-)}(\mathbf{0})\chi_p^{(+)}(\mathbf{0})=W_{0}(\eta_p)\,C_{0}^2(\eta_p)  e^{2i\sigma_0}.
\end{eqnarray}
According to the diagrams in second line of Fig.~\ref{fig:2}, The $S$-wave up-to-NLO full propagator is given by
\begin{eqnarray}{\label{eq:a5}}
D^{[\xi]}(E,\textbf{0})&=&\frac{\eta^{[\xi]}}{E-\mathrm{\Delta}^{[\xi]}-\eta^{[\xi]}g^{[\xi]^2}J_0(E)} \Big[\underbrace{\,1_{_{_{_{_{_{_{_{_{}}}}}}}}}}_{\mathrm{LO}}-\underbrace{\frac{\eta^{[\xi]}h^{[\xi]}E^2}{E-\mathrm{\Delta}^{[\xi]}-\eta^{[\xi]}g^{[\xi]^2}J_0(E)}}_{\mathrm{NLO}~ \mathrm{corection}}\Big],\quad\quad\quad
\end{eqnarray}
where the fully dressed bubble $J_{0}$, which is described the propagation of the particles from initially zero separation and back to zero separation, is written as
\begin{eqnarray}{\label{eq:a6}}
J_{0}(E)&=&\lim_{\mathbf{r}^{\prime},\mathbf{r}\rightarrow \mathbf{0}}\langle\mathbf{r}^{\prime}|G_{C}^{(+)}{(E)}|\mathbf{r}\rangle\nonumber\\
&=&2\mu\int\frac{d^{3}q}{(2\pi)^{3}}\frac{\chi_{q}^{(+)}(\mathbf{0})\chi_{q}^{*(+)}(\mathbf{0})}{2\mu E-q^{2}+i\epsilon}
\nonumber\\
&=&2\mu\!\int \!\frac{d^{3}q}{(2\pi)^{3}}\frac{2\pi\eta_q}{e^{2\pi\eta(q)}-1}\,\frac{1}{p^2-q^{2}+i\epsilon}\nonumber\\
&=&\underbrace{2\mu\!\int\! \frac{d^{3}q}{(2\pi)^{3}}\frac{2\pi\eta_q}{e^{2\pi\eta_q}-1}\,\frac{1}{q^2}\frac{p^2}{p^2-q^{2}+i\epsilon}}_{J^{fin}_{0}}\nonumber\\
&& \underbrace{-2\mu\!\int\! \frac{d^{3}q}{(2\pi)^{3}}\frac{2\pi\eta_q}{e^{2\pi\eta_q}-1}\,\frac{1}{q^2}}_{J^{div}_{0}}.\qquad\qquad
\end{eqnarray}
Calculation of the finite part of the $S$-wave Coulomb bubble leads to~\cite {33}
\begin{eqnarray}{\label{eq:a7}}
J^{fin}_0=-\frac{\mu }{\pi}k_CW_0(\eta_p)H(\eta_p)=-\frac{\mu }{2\pi}H_0(\eta_p),
\end{eqnarray}
and taking into account the power divergence subtraction (PDS) regularization scheme, the momentum independent divergent part is obtained as~\cite {33}
\begin{eqnarray}\label{eq:a8}
J^{div}_0\!=\!-\frac{\mu}{2\pi}\bigg{\{}\!\frac{\kappa}{D-3}\!+\!2k_C\! \bigg[\!\frac{1}{D-4}\!-\!\textrm{ln}\big(\frac{\kappa\sqrt{\pi}}{2k_C}\big)\!-\!1\!+\!\frac{3}{2}C_E\!\bigg]\!\bigg{\}},\nonumber\\\!\!\!\!\!\!\!\!\!\!\!\!
\end{eqnarray}
with $D$ the dimensionality of spacetime, $\kappa$ the renormalization mass scale and $C_E$ Euler-Masheroni constant.
Instead of PDS regularization scheme we can use a simple momentum cutoff $\mathrm{\Lambda}$ to make the divergent integral $J^{div}_0$ finite.
It then becomes~\cite {33}
\begin{eqnarray}\label{eq:a8}
J^{div}_0&\!=\!&-\frac{2\mu}{\pi}\!\int_{0}^{\mathrm{\Lambda}}\!dq \frac{\eta_q}{e^{2\pi\eta_q}-1}\nonumber\\
&\!=\!&-\frac{2\mu k_C}{\pi}\!\int_{ \frac{2\pi k_C}{\mathrm{\Lambda}}}^{\infty}\,\! \frac{dx}{x(e^{x}-1)}\nonumber\\
&\!=\!&-\frac{2\mu k_C}{\pi}\Bigg\{\!\int_{0}^{\infty}\! \frac{dx}{x(e^{x}-1)}\!-\!\int_{0}^{ \frac{2\pi k_C}{\mathrm{\Lambda}}}\frac{dx}{x(e^{x}-1)}\Bigg\}\! \nonumber\\
&\!=\!&-\frac{2\mu k_C}{\pi}\Bigg\{\!\mathrm{\Gamma}(0)\zeta(0)\!-\!\int_{0}^{ \frac{2\pi k_C}{\mathrm{\Lambda}}}\!\!dx\bigg(\!\frac{1}{x^2}-\frac{1}{2x}+\mathcal{O}\,(x^0)\!\bigg)\!\!\Bigg\}\!\nonumber\\
&\!=\!&-\frac{2\mu k_C}{\pi}\Big(\frac{1}{2}C_E+\frac{\mathrm{\Lambda}}{2\pi k_C}-\frac{1}{2}\ln\frac{\mathrm{\Lambda}}{k_C}+\mathcal{O}\,(\frac{2\pi k_C}{\mathrm{\Lambda}})\!\Big),\nonumber\\
\end{eqnarray}
where in the second line we use changing integral variable $x=2\pi\eta_q$, and in the last line we use
\begin{eqnarray}\label{eq:a10}
 \mathrm{\Gamma}(0)&=&\lim_{\epsilon \rightarrow0}\big(\frac{1}{\epsilon}-C_E\big),\\
 \zeta(0)&=&\lim_{\epsilon \rightarrow0}\Big(-\frac{1}{2}(1+\epsilon \ln 2\pi)+\mathcal{O}\,(\epsilon^2)\Big).
\end{eqnarray}
Thus, the up-to-NLO EFT scattering amplitude of Eq. (\ref{eq:a2}) is rewritten
\begin{eqnarray}\label{eq:a10}
 \!\!T^{[\xi]}_{CS}\!&=&\!-\frac{2\pi}{\mu}\frac{C_{0}^2(\eta_p)W_{0}(\eta_p)}{(\frac{2\pi\mathrm{\Delta}^{[\xi]}}{\eta^{[\xi]}g^{[\xi]^2}\!\mu}\!+\!\frac{2\pi}{\mu}J^{div}_0)\!-\!\frac{1}{2}(\frac{2\pi }{\eta^{[\xi]}g^{[\xi]^2}\!\mu^2})p^2\!-\!H_0(\eta_p)}  \nonumber\\&&\times\Big[\underbrace{\,1_{_{_{_{_{_{_{_{_{_{_{_{_{_{}}}}}}}}}}}}}}}_{\mathrm{\!\!LO}}\!\!\!\!+\!\underbrace{\frac{1}{4}\frac{(\frac{2\pi h^{[\xi]}}{g^{[\xi]^2}\!\mu^3})}{(\frac{2\pi\mathrm{\Delta}^{[\xi]}}{\eta^{[\xi]}g^{[\xi]^2}\mu}\!+\!\frac{2\pi}{\mu}J^{div}_0)\!-\!\frac{1}{2}(\frac{2\pi }{\eta^{[\xi]}g^{[\xi]^2}\!\mu^2})p^2\!-\!H_0(\eta_p)}p^4\,\,}_{\mathrm{NLO}~\mathrm{corection}}\!\!\Big]\!.\nonumber\\
\end{eqnarray}
Regardless of which renormalization scheme we use to calculate the divergent integral $J^{div}_0$, this momentum independent divergence part is absorbed by the parameter $\mathrm{\Delta}^{[\xi]}$ via introducing the renormalized
parameter $\mathrm{\Delta}^{[\xi]}_R$ as~\cite {Higa-Hammer}
\begin{equation}
\mathrm{\Delta}^{[\xi]}_R=\mathrm{\Delta}^{[\xi]}+\eta^{[\xi]}g^{[\xi]^2}J^{div}_0.
\end{equation}
Finally, the up-to-NLO scattering amplitude for $\xi\!=\!{^3\!S_1}$ partial wave is expressed as
\begin{eqnarray}\label{eq:a100}
 T^{[\xi]}_{CS}&=&\frac{2\pi}{\mu}\frac{C_{0}^2(\eta_p)W_{0}(\eta_p)}{\frac{2\pi\mathrm{\Delta}_R^{[\xi]}}{\eta^{[\xi]}g^{[\xi]^2}\!\mu}-\frac{1}{2}(\frac{2\pi }{\eta^{[\xi]}g^{[\xi]^2}\!\mu^2})p^2\!-\!H_0(\eta_p)}\nonumber\\&&\times  \Big[\underbrace{\,\,1_{_{_{_{_{_{_{_{_{_{_{_{_{_{_{_{}}}}}}}}}}}}}}}}}_{\mathrm{LO}}+\!\underbrace{\frac{1}{4}\frac{(\frac{2\pi h^{[\xi]}}{g^{[\xi]^2}\!\mu^3})}{\frac{2\pi\mathrm{\Delta}_R^{[\xi]}}{\eta^{[\xi]}g^{[\xi]^2}\!\mu}-\frac{1}{2}(\frac{2\pi }{\eta^{[\xi]}g^{[\xi]^2}\!\mu^2})p^2\!-\!H_0(\eta_p)}p^4\,\,}_{\mathrm{NLO}~\mathrm{corection}}\!\!\Big].\nonumber\\
\end{eqnarray}

\subsection*{  $P-$wave channels}{\label{subsec:5}}
The up-to-NLO Lagrangian for the strong interaction in the $\xi=$${^3\!P_0}$ channel of the $d-\alpha$ system can be written as
\begin{eqnarray}{\label{eq:a15}}
\mathcal{L}^{[\xi]}&=&\phi^{\dagger}(i\partial _0+\frac{\nabla ^2}{2m_\alpha})\phi+d_i^{\dagger}(i\partial _0+\frac{\nabla ^2}{2m_d})d_i
\nonumber\\
&&+\,\eta^{[{\xi}]}t^\dagger\Big[i\partial _0
+\frac{\nabla ^2}{2m_t}-\mathrm{\Delta}^{[{\xi}]}\Big]t +h^{[{\xi}]}t^{\dagger}\Big[i\partial _0+\frac{\nabla ^2}{2m_t}\Big]^2t\nonumber\\
&& +\sqrt{3}\,g^{[{\xi}]}\Big[t^{\dagger}(\phi \mathcal{P} _i d_i)+h.c.\Big],
\end{eqnarray}
where $t$  is the  scaler auxiliary field of the $^3\!P_0$ dimeron. According to the Feynman diagrams of Fig.~\ref{fig:2}
we have
\begin{eqnarray}\label{eq:a16}
-i3T^{[\xi]}_{CS}P_1(\hat{\textbf{p}}'\cdot \hat{\textbf{p}})e^{2i\sigma_1}&=&3(-ig^{[\xi]})^2[\mathcal{P}^*_j\chi_{p'}^{*(-)}(\mathbf{0})] \varepsilon_j^{d*} iD^{[\xi]}(E,\textbf{0})\,\varepsilon_i^{d}\,[\mathcal{P}_i\chi_p^{(+)}(\mathbf{0})]\nonumber\\
&=&-3ig^{[\xi]^2}D^{[\xi]}(E,\textbf{0}) \varepsilon_j^{d*}\,\varepsilon_i^{d*}[\nabla_j\chi_{p'}^{*(-)}(\mathbf{0})][\nabla_i\chi_p^{(+)}(\mathbf{0})]\nonumber\\
&=&-ig^{[\xi]^2}D^{[\xi]}(E,\textbf{0})  C_0^2(\eta_p) W_1(\eta_p) P_1(\hat{\textbf{p}}'\cdot \hat{\textbf{p}}) e^{2i\sigma_1},\quad\quad\quad
\end{eqnarray}
where in the last line, the following relation is used
\begin{eqnarray}{\label{eq:a17}}
[\nabla_i \chi_{p'}^{*(-)}(\mathbf{0})][\nabla_i \chi_p^{(+)}(\mathbf{0})]&=&C^2_0(\eta_p)\, p'_ip_{i}\,(1+\eta_p^2)e^{2i\sigma_1}\nonumber\\
&=&C^2_0(\eta_p)\, W_1(\eta_p) P_1(\hat{\textbf{p}}'\cdot \hat{\textbf{p}})e^{2i\sigma_1}.\nonumber\\
\end{eqnarray}
The up-to-NLO strong interaction Lagrangian in the $\xi=$${^3\!P_1}$ channel is introduced as
\begin{eqnarray}{\label{eq:a18}}
\mathcal{L}^{[\xi]}&=&\phi^{\dagger}(i\partial _0+\frac{\nabla ^2}{2m_\alpha})\phi+d_i^{\dagger}(i\partial _0+\frac{\nabla ^2}{2m_d})d_i\nonumber\\
&& +\,\eta^{[\xi]}t_i^\dagger\Big[i\partial _0+\frac{\nabla ^2}{2m_t}-\mathrm{\Delta}^{[\xi]}\Big]t_i+h^{[\xi]}t_i^{\dagger}\Big[i\partial _0+\frac{\nabla ^2}{2m_t}\Big]^2t_i\nonumber\\
&& +\sqrt{\frac{3}{2}}\epsilon _{kji}\, g^{[\xi]}\Big[t_k^{\dagger}(\phi \mathcal{P}_j d_i)+h.c.\Big],
\end{eqnarray}
where $t_i$  denotes the vector field of the $^3\!P_1$ dimeron. So, the scattering amplitude in the $^3\!P_1$ channel is written as
\begin{eqnarray}{\label{eq:a19}}
-i3T^{[\xi]}_{CS}P_1(\hat{\textbf{p}}'\cdot \hat{\textbf{p}})e^{2i\sigma_1}&=&\frac{3}{2}(-ig^{[\xi]})^2\, [\mathcal{P}^*_m\chi_{p'}^{*(-)}(\mathbf{0})] \epsilon_{lmj} \varepsilon_j^{d*}\,\varepsilon^{t}_l iD^{[\xi]}(E,\textbf{0})\epsilon_{ksi}\varepsilon_{k}^{t*}\,\varepsilon_i^{d}\,[\mathcal{P}_s\chi_p^{(+)}(\mathbf{0})]\nonumber\\%&=-\frac{1}{2}\delta_{lk}\delta_{ij}\,ig^{[\xi]^2}\, D^{[n,\xi]}(E,\textbf{0})\,\epsilon_{lmj}\,\epsilon_{ksi}\,[\nabla_m\chi_{p'}^{*(-)}(\mathbf{0})]\,[\nabla_s\chi_p^{(+)}(\mathbf{0})]\nonumber\\
&=&-\frac{1}{2}\,ig^{[\xi]^2}\, D^{[\xi]}(E,\textbf{0})\,\epsilon_{kmi}\,\epsilon_{ksi} [\nabla_m\chi_{p'}^{*(-)}(\mathbf{0})]\,[\nabla_s\chi_p^{(+)}(\mathbf{0})]\nonumber\\
%&=-ig^{[\xi]^2}\, D^{[n,\xi]}(E,\textbf{0})\,[\nabla_i\chi_{p'}^{*(-)}(\mathbf{0})]\,[\nabla_i\chi_p^{(+)}(\mathbf{0})]\nonumber\\
&=&-ig^{[\xi]^2}D^{[\xi]}(E,\textbf{0})  C_0^2(\eta_p)  W_1(\eta_p) P_1(\hat{\textbf{p}}'\cdot \hat{\textbf{p}}) e^{2i\sigma_1},\quad
\end{eqnarray}
with $\varepsilon_i^{t}$ as the polarization vector of the $^3\!P_1$ dimeron auxiliary field. Also, the strong interaction Lagrangian for the $d-\alpha$ system in the $\xi={^3\!P_2}$ channel can be written as
\begin{eqnarray}{\label{eq:a20}}
\mathcal{L}^{[\xi]}&=&\phi^{\dagger}(i\partial _0+\frac{\nabla ^2}{2m_\alpha})\phi+d_i^{\dagger}(i\partial _0+\frac{\nabla ^2}{2m_d})d_i
\nonumber\\
&& +\,\eta^{[\xi]}t_{ij}^\dagger\!\Big[i\partial _0\!+\!\frac{\nabla ^2}{2m_t}\!-\!\mathrm{\Delta}^{[\xi]}\!\Big]t_{ij}\!+\!h^{[\xi]}t_{ij}^\dagger\!\Big[i\partial _0+\frac{\nabla ^2}{2m_t}\Big]^2\!t_{ij}\nonumber\\&& +\frac{3}{\sqrt{5}} g^{[\xi]}\Big[\!t_{ij}^{\dagger}(\phi \mathcal{P} _j d_i)+h.c.\Big],
\end{eqnarray}
where $t_{ij}$ is the auxiliary tensor field of the~$^3\!P_2$ dimeron. Therefore, the scattering amplitude in the $^3\!P_2$ channel is obtained as
\begin{eqnarray}{\label{eq:a21}}
-3iT^{[\xi]}_{CS}P_1(\hat{\textbf{p}}'\cdot \hat{\textbf{p}})e^{2i\sigma_1}&=&\frac{9}{5}(-ig^{[\xi]})^2\,[\mathcal{P}^*_m\chi_{p'}^{*(-)}(\mathbf{0})]  \varepsilon_j^{d*}\,\varepsilon_{jm}^t iD^{[\xi]}(E,\textbf{0}) \varepsilon_{si}^{t*}\,\varepsilon_i^{d}\,[\mathcal{P}_s\chi_p^{(+)}(\mathbf{0})]\nonumber\\%&=-\frac{9}{5}ig^{[\xi]^2}\, D^{[\xi]}(E,\textbf{0})\,  \varepsilon_j^{d*}\,\varepsilon_{mj}^t\,\varepsilon_{is}^{t*}\,\varepsilon_i^{d}[\nabla_m\chi_{p'}^{*(-)}(\mathbf{0})]\,[\nabla_s\chi_p^{(+)}(\mathbf{0})]\nonumber\\
%&=-ig^{[\xi]^2}\, D^{[\xi]}(E,\textbf{0})\,  [\nabla_i\chi_{p'}^{*(-)}(\mathbf{0})]\,[\nabla_i\chi_p^{(+)}(\mathbf{0})]\nonumber\\
&=&-ig^{[\xi]^2}\!D^{[\xi]}(E,\textbf{0}) C_0^2(\eta_p)  W_1(\eta_p) P_1(\hat{\textbf{p}}'\cdot \hat{\textbf{p}}) e^{2i\sigma_1},\qquad\qquad
\end{eqnarray}
with $\varepsilon_{ij}$ as the polarization tensor of the $^3\!P_2$ dimeron auxiliary field which satisfies the expression
\begin{equation}{\label{eq:a22}}
\varepsilon_{jm}^t\,\varepsilon_{si}^{t*}=\frac{1}{2}(\delta_{js}\delta_{mi}+\delta_{ji}\delta_{ms}-\frac{2}{3}\delta_{jm}\delta_{si}).
\end{equation}
The up-to-NLO full propagator for the $^3\!P_0$, $^3\!P_1$ and $^3\!P_2$ channels is given by
\begin{eqnarray}{\label{eq:a23}}
D^{[\xi]}(E,\textbf{0})&=&\frac{\eta^{[\xi]}}{E-\mathrm{\Delta}^{[\xi]}-\frac{1}{3}\eta^{[\xi]}g^{[\xi]^2}J_1(E)}\Big[\underbrace{1_{_{_{_{_{_{_{_{_{_{_{}}}}}}}}}}}}_{\mathrm{LO}}-\underbrace{\frac{\eta^{[\xi]}h^{[\xi]}E^2}{E\!-\mathrm{\Delta}^{[\xi]}\!-\frac{1}{3}\eta^{[\xi]}g^{[\xi]^2}J_1(E)}}_{\mathrm{NLO~corection}}\!\Big]\!.\quad\quad\quad
\end{eqnarray}
The function $J_1(E)$ is given by
\begin{eqnarray}\label{eq:555}
J_1(E)&\!=&2\mu \!\!\int\! \!\frac{d^3 q}{(2 \pi )^3}\frac{[\nabla_i\chi _{{q}}^{(+)}(\mathbf{0})] [\nabla_i\chi _{{q}}^{*(+)}(\mathbf{0})]}{2\mu E -{q}^2+i\epsilon }
\nonumber\\
&=&2\mu \!\!\int\! \!\frac{d^3 q}{(2 \pi )^3}\frac{q^2 +k_C^2}{p^2-{q}^2+i\epsilon }\,\frac{2 \pi \eta_q }{e^{2 \pi \eta_q}-1}\nonumber\\
&=&2\mu \!\!\int\!\! \frac{d^3 q}{(2 \pi )^3}\frac{q^2 }{p^2-{q}^2+i\epsilon }\,\frac{2 \pi \eta_q }{e^{2 \pi \eta_q}-1}\!+\!k_C^2J_0(E)\nonumber\\
&=&2\mu \!\!\int \!\!\frac{d^3 q}{(2 \pi )^3}\frac{q^2-p^2 }{p^2-{q}^2+i\epsilon }\,\frac{2 \pi \eta_q }{e^{2 \pi \eta_q}-1}\!+\!(p^2\!+\!k_C^2)J_0(E)\nonumber\\
&=&W_1(\eta_p)J_0(E)\underbrace{-2\mu \int \frac{d^3 q}{(2 \pi )^3}\,\frac{2 \pi \eta_q }{e^{2 \pi \eta_q}-1}}_{J}.\qquad
\end{eqnarray}
In the second line of Eq.~(\ref{eq:555}) we use
\begin{eqnarray}
[\nabla_i \chi_{q}^{(+)}(\mathbf{0})][\nabla_i \chi_q^{*(+)}(\mathbf{0})]\!=\!C^2_0(\eta_q)W_1(\eta_q).
\end{eqnarray}
The integral $J$ is divergent and independent of the external momentum $p$. According to the PDS regularization scheme it takes the form~\cite{Higa-Hammer}
\begin{eqnarray}
J=-4\pi \mu k_C^2  \Big( k_C\zeta'(-2)+\frac{\kappa}{24}\Big),
\end{eqnarray}
where $\zeta'$ is derivative of the Riemann zeta function and $\zeta'(-2)\!\approx\!-0.0304$.
If we use the cutoff regularization scheme the integral $J$ takes the form
\begin{eqnarray}\label{eq:a8}
J&\!=\!&-\frac{2\mu}{\pi}\!\int_{0}^{\mathrm{\Lambda}}\!dq q^2 \frac{\eta_q}{e^{2\pi\eta_q}-1}\nonumber\\
&\!=\!&-8\pi \mu k^3_C\!\int_{\frac{2\pi k_C}{\mathrm{\Lambda}}}^{\infty}\!\, \frac{dx}{x^3(e^{x}-1)}
\nonumber\\
&\!=\!&-8\pi \mu k^3_C\Bigg\{\!\int_{0}^{\infty}\! \frac{dx}{x^3(e^{x}-1)}-\int_{0}^{\frac{2\pi k_C}{\mathrm{\Lambda}}}\frac{dx}{x^3(e^{x}-1)}\Bigg\}\! \nonumber\\
&\!=\!&-8\pi\mu  k^3_C\Bigg\{\mathrm{\Gamma}(-2)\zeta(-2)-\int_{0}^{ \frac{2\pi k_C}{\mathrm{\Lambda}}}\!\!dx\bigg(\!\frac{1}{x^4}-\frac{1}{2x^3}+\frac{1}{12x^2}+\mathcal{O}\,(x^0)\!\bigg)\!\!\Bigg\}\!\nonumber\\
&\!=\!&-8\pi \mu k^3_C\bigg\{ 2\pi^2C_E \,\zeta'(-2) \!+\!\frac{1}{3}\Big(\frac{\mathrm{\Lambda}}{2\pi k_C }\Big)^{\!3}\!-\!\frac{1}{4}\Big(\frac{ \mathrm{\Lambda}}{2\pi k_C}\Big)^{\!2}\!+\!\frac{1}{12}\Big(\frac{\mathrm{\Lambda}}{2\pi k_C}\Big)+\mathcal{O}(\frac{2\pi k_C}{\mathrm{\Lambda}})\!\bigg\}
\end{eqnarray}
where in the second line we use $x=2\pi \eta_q$. Thus, $J_1$ can be divided as $J_1\!=\!J_1^{fin}\!+\!J_1^{div}$ with
\begin{eqnarray}
J^{fin}_1&=&W_1(\eta_p)J^{fin}_0=-\frac{\mu }{2\pi}H_1(\eta_p),\\
J^{div}_1&=&W_1(\eta_p)J^{div}_0+J=p^2J^{div}_0+(k_C^2J^{div}_0+J).\qquad\quad
\end{eqnarray}
Consequently, the up-to-NLO EFT scattering amplitude of Eqs. (\ref{eq:a16}), (\ref{eq:a19}) and (\ref{eq:a21}) is rewritten as
\begin{eqnarray}\label{eq:a28}
 T^{[\xi]}_{CS}&=&-\frac{2\pi}{\mu}\frac{C_{0}^2(\eta_p)W_{1}(\eta_p)}{(\frac{6\pi\mathrm{\Delta}^{[\xi]}}{\eta^{[\xi]}g^{[\xi]^2}\mu}\!+\!\frac{2\pi}{\mu}(k_C^2J^{div}_0+J))\!-\!\frac{1}{2}(\frac{6\pi }{\eta^{[\xi]}g^{[\xi]^2}\mu^2}+\frac{2\pi}{\mu}J^{div}_0)p^2\!-\!H_1(\eta_p)}  \nonumber\\&&\!\!\!\!\times\Big[\underbrace{1_{_{_{_{_{_{_{_{_{_{_{_{_{}}}}}}}}}}}}}}_{\mathrm{LO}}+\!\underbrace{\frac{1}{4}\frac{(\frac{6\pi h^{[\xi]}}{g^{[\xi]^2}\mu^3})}{(\frac{6\pi\mathrm{\Delta}^{[\xi]}}{\eta^{[\xi]}g^{[\xi]^2}\mu}\!+\!\frac{2\pi}{\mu}(k_C^2J^{div}_0+J))\!-\!\frac{1}{2}(\frac{6\pi }{\eta^{[\xi]}g^{[\xi]^2}\mu^2}+\frac{2\pi}{\mu}J^{div}_0)p^2\!-\!H_1(\eta_p)}p^4\,\,}_{\mathrm{NLO}~\mathrm{corection}}\!\!\!\Big]
\end{eqnarray}
\noindent{\!The function $J_1^{div}$ has two divergences, momentum independent and momentum-squared.
Regardless of PDS or cutoff renormalization scheme are used to calculate
the divergent integrals $J_0^{div}$ and $J$, these momentum independent and
momentum-squared divergence parts are absorbed by the parameters $\mathrm{\Delta}^{[\xi]}$, $g^{[\xi]}$ and $h^{[\xi]}$
via introducing the renormalized parameters $\mathrm{\Delta}_R^{[\xi]}$, $g_R^{[\xi]}$ and $h_R^{[\xi]}$ as}
\begin{align}
\mathrm{\Delta}^{[\xi]}_R&=\frac{\mathrm{\Delta}^{[\xi]}+\frac{1}{3}\eta^{[\xi]}g^{[\xi]^2} (k_C^2J^{div}_0+J)}{1+\frac{1}{3}\eta^{[\xi]}g^{[\xi]^2}\mu J^{div}_0},\\
\frac{1}{g_R^{[\xi]^2}}&=\frac{1}{g^{[\xi]^2}}+\frac{1}{3}\eta^{[\xi]}\mu J^{div}_0,\\
h_R^{[\xi]}&=\frac{h^{[\xi]}}{1+\frac{1}{3}\eta^{[\xi]}g^{[\xi]^2}\mu J^{div}_0}.
\end{align}
Finally, the up-to-NLO Coulomb-subtracted EFT scattering amplitude for $^3\!P_0$, $^3\!P_1$ and $^3\!P_2$ channels are obtained
\begin{eqnarray}{\label{eq:a32}}
T^{[\xi]}_{CS}&=&-\frac{2\pi}{\mu}\frac{ C_0^2(\eta_p)W_1(\eta_p) }{\frac{6\pi\mathrm{\Delta}_{\!R}^{[\xi]}}{\eta^{[\xi]}g_{\!R}^{[\xi]^2}\mu}\!-\!\frac{1}{2}(\frac{6\pi }{\eta^{[\xi]}g_{\!R}^{[\xi]^2}\mu^2})p^2\!-\!H_1(\eta_p)} \Big[\underbrace{\,\,1_{_{_{_{_{_{_{_{_{_{_{_{_{_{_{_{_{_{}}}}}}}}}}}}}}}}}}}_{\mathrm{LO}}+\underbrace{\frac{1}{4}\frac{(\frac{6\pi h_{\!R}^{[\xi]}}{g_{\!R}^{[\xi]^2}\mu^3})}{\frac{6\pi\mathrm{\Delta}_{\!R}^{[\xi]}}{\eta^{[\xi]}g_{\!R}^{[\xi]^2}\mu}\!-\!\frac{1}{2}(\frac{6\pi }{\eta^{[\xi]}g_{\!R}^{[\xi]^2}\mu^2})p^2\!-\!H_1(\eta_p)}p^4\,}_{\mathrm{NLO~corection}}\!\!\Big]\!,\nonumber\\
\end{eqnarray}

\subsection*{ $D-$wave channels}{\label{subsec:6}}
The Lagrangian for the strong $d-\alpha$ interaction in the $\xi=$${^3D_1}$ channel is written as
\begin{eqnarray}{\label{eq:a555}}
\mathcal{L}^{[\xi]}&\!=&\!\phi^{\dagger}(i\partial _0+\frac{\nabla ^2}{2m_\alpha})\phi+d_i^{\dagger}(i\partial _0+\frac{\nabla ^2}{2m_d})d_i
\nonumber\\&&\! +\,\tilde{t}_{i}^{\,\dagger}\!\Big[\eta^{[\xi]}(i\partial _0\!+\!\frac{\nabla ^2}{2m_t}\!-\!\mathrm{\Delta}^{[\xi]})\Big]\tilde{t}_{i}\! +\!\tilde{t}_{i}^{\,\dagger}\!\Big[h^{[\xi]}(i\partial _0\!+\!\frac{\nabla ^2}{2m_t})^2\Big]\tilde{t}_{i}\nonumber\\
&&\! +\frac{3}{\sqrt{2}}\,g^{[\xi]}\Big[\tilde{t}_{j}^{\,\dagger}(\phi \,\tau_{ji} d_i)+h.c.\Big],
\end{eqnarray}
where $\tilde{t}_i$ is the vector field of the $\!{^3\!D_1}$ dimeron. Using the Lagrangian~(\ref{eq:a555}), the Coulomb-subtracted amplitude in ${^3\!D_1}$ partial wave is evaluated by
\begin{eqnarray}{\label{eq:a36}}
-i5T^{[\xi]}_{CS}P_2(\hat{\textbf{p}}'\cdot\hat{\textbf{p}})e^{2i\sigma_2}&=&\frac{9}{2}(-ig^{[\xi]})^2[\tau^*_{jl}\chi_{p'}^{*(-)}\!(\mathbf{0})] \varepsilon_j^{d*}\varepsilon^{\tilde{t}}_l iD^{[\xi]}(E,\textbf{0})\varepsilon_{k}^{\tilde{t}*}\varepsilon_i^{d}[\tau_{ki}\chi_p^{(+)}\!(\mathbf{0})]\nonumber\\%&=-\frac{3}{2}\delta_{lk}\,\delta_{ij}\,ig^{[\xi]^2}\, D^{[n,\xi]}(E,\textbf{0})\,[\tau_{lj}\chi_{p'}^{*(-)}(\mathbf{0})]\,[\tau_{ki}\chi_p^{(+)}(\mathbf{0})]\nonumber\\
&=&-\frac{3}{2}ig^{[\xi]^2} D^{[\xi]}(E,\textbf{0}) [\tau^*_{ki}\chi_{p'}^{*(-)}(\mathbf{0})]\,[\tau_{ki}\chi_p^{(+)}(\mathbf{0})]\nonumber\\&=&-ig^{[\xi]^2} D^{[\xi]}(E,\textbf{0})\,C_0^2(\eta_p)W_2(p)  P_2(\hat{\textbf{p}}'\cdot\hat{\textbf{p}})e^{2i\sigma_2},\quad
\end{eqnarray}{\label{eq:a29}}
\noindent{\!\!where $\varepsilon^{\tilde{t}}_i$ is the vector auxiliary field of the $\!{^3\!D_1}$ dimeron and in the last equality we use
\begin{eqnarray}
[\tau^*_{ki}\chi_{p'}^{*(-)}\!(\mathbf{0})][\tau_{ki}\chi_p^{(+)}\!(\mathbf{0})]&=&\frac{1}{4}(p'_k p_k \,p'_ip_i-\frac{1}{3}p'^{2}p^2\delta_{ki}) C^2_0(\eta_p)(1+\eta_p^2)(4+\eta_p^2)e^{2i\sigma_2}\nonumber\\
&=&\frac{1}{6}C^2_0(\eta_p)p^4(1+\eta_p^2)(4+\eta_p^2)P_2(\hat{\textbf{p}}'\cdot\hat{\textbf{p}})e^{2i\sigma_2}\nonumber\\&=&\frac{2}{3}W_2(p)P_2(\hat{\textbf{p}}'\cdot\hat{\textbf{p}})e^{2i\sigma_2}.\qquad\quad\qquad\qquad\qquad\qquad\qquad
\end{eqnarray}
In order to calculate the Coulomb-subtracted EFT amplitude of $d-\alpha$ scattering in the $\xi=$${^3\!D_2}$ channel, we introduce the strong interaction in this channel using the Lagrangian
\begin{eqnarray}{\label{eq:a30}}
\mathcal{L}^{[\xi]}&=&\phi^{\dagger}(i\partial _0+\frac{\nabla ^2}{2m_\alpha})\phi+d_i^{\dagger}(i\partial _0+\frac{\nabla ^2}{2m_d})d_i\nonumber\\
&&
+\,\tilde{t}_{ij}^{\,\dagger}\Big[\eta^{[\xi]}(i\partial _0+\frac{\mathrm{\nabla} ^2}{2m_t}-\mathrm{\Delta}^{[\xi]})+h^{[\xi]} (i\partial _0+\frac{\nabla ^2}{2m_t})^2\Big]\tilde{t}_{ij}\nonumber\\&&+ \sqrt{\frac{3}{2}}\epsilon _{lji} \,g^{[\xi]}[\tilde{t}_{kl}^{\,\dagger}(\phi \,\tau_{kj} d_i)+h.c.],
\end{eqnarray}
with $\tilde{t}_{ij}$ as the ${^3\!D_2}$ tensor auxiliary field. So, we have
\begin{eqnarray}{\label{eq:a31}}
-i5T^{[\xi]}_{CS}P_2(\hat{\textbf{p}}'\cdot\hat{\textbf{p}})e^{2i\sigma_2}&=&\frac{3}{2}(\!-ig^{[\xi]})^2[\tau^*_{mn} \chi_{p'}^{*(-)}(\mathbf{0})]   \varepsilon_j^{*d}\epsilon _{snj}\varepsilon_{ms}^{\tilde{t}}\nonumber\\
&&\times iD^{[\xi]}(E,\textbf{0}) \varepsilon_{kp}^{*\tilde{t}}\,\epsilon _{pli} \,\varepsilon_i^{d}\,[\tau_{kl}\chi_p^{(+)}(\mathbf{0})]\nonumber\\&=&-\frac{3}{2}ig^{[\xi]^2}D^{[\xi]}(E,\textbf{0})  [\tau^*_{ki}\chi_{p'}^{*(-)}(\mathbf{0})][\tau_{ki}\chi_p^{(+)}(\mathbf{0})]\nonumber\\&=&-ig^{[\xi]^2} D^{[\xi]}(E,\textbf{0})\,C_0^2(\eta_p) W_2(p) P_2(\hat{\textbf{p}}'\cdot\hat{\textbf{p}})e^{2i\sigma_2},\qquad\quad
\end{eqnarray}
Also, the strong interaction Lagrangian of the $d-\alpha$ system in the $\xi\!=$${^3\!D_3}$ channel  can be described as
\begin{eqnarray}{\label{eq:a32}}
\mathcal{L}^{[\xi]}&=&\phi^{\dagger}(i\partial _0+\frac{\nabla ^2}{2m_\alpha})\phi+d_i^{\dagger}(i\partial _0+\frac{\nabla ^2}{2m_d})d_i
+\sqrt{\frac{45}{8}} g^{[\xi]}[\tilde{t}_{ijk}^{\,\dagger}(\phi\tau_{ij} d_k)+h.c.]\nonumber\\
&& +\,\tilde{t}_{ijk}^{\,\dagger}\Big[\eta^{[\xi]}(i\partial _0+\frac{\nabla ^2}{2m_t}-\mathrm{\Delta}^{[\xi]})\Big]\tilde{t}_{ijk}+\,h^{[\xi]} (i\partial _0+\frac{\nabla ^2}{2m_t})^2\Big]\tilde{t}_{ijk},\qquad\qquad\qquad\qquad
\end{eqnarray}
where $t_{ijk}$ indicates the auxiliary tensor field of the $^3\!D_3$ dimeron. According to the  Feynman diagram of Fig.~2, we have
\begin{eqnarray}{\label{eq:a33}}
-i5T^{[\xi]}_{CS}P_2(\hat{\textbf{p}}'\cdot\hat{\textbf{p}})e^{2i\sigma_2}&=&\frac{45}{8}(-ig^{[\xi]})^2 [\tau^*_{kl}\chi_{p'}^{(-)*}(\mathbf{0})] \varepsilon^{*d}_{j}\varepsilon_i^{d} iD^{[\xi]}(E,\textbf{0}) \,\varepsilon^{\tilde{t}}_{klj} \varepsilon_{mni}^{*\tilde{t}}\,[\tau_{mn}\chi_p^{(+)}(\mathbf{0})]\nonumber\\&=&-ig^{[\xi]^2} D^{[\xi]}(E,\textbf{0})\,C_0^2(\eta_p) W_2(p) P_2(\hat{\textbf{p}}'\cdot\hat{\textbf{p}})e^{2i\sigma_2},\qquad\quad\\\nonumber
\end{eqnarray}
where $\varepsilon_{ijk}$ denotes the tensor polarization of $^3\!D_3$ auxiliary field which satisfies the following relation
\begin{eqnarray}
\varepsilon^{\tilde{t}}_{klj} \varepsilon_{mni}^{*\tilde{t}}&=&\frac{1}{6}\bigg[\!-\frac{2}{5}\bigg{\{}\delta_{mn}(\delta_{ij}\delta_{kl}+\delta_{ik}\delta_{jl}+\delta_{il}\delta_{jk})+\,(m\! \leftrightarrow \!l )+(n \!\leftrightarrow \!l )\bigg{\}}\nonumber\\
&&\,\,\,\,\,\,\,\,+\,(\delta_{il}\delta_{jm}\delta_{kn} +\delta_{il}\delta_{jn}\delta_{km})+\,(i \rightarrow\! j\rightarrow \!k\rightarrow\! i)+(i \rightarrow k\rightarrow j \rightarrow i)\bigg].\nonumber\\
\end{eqnarray}
The full propagator for $D$ waves is expressed by
\begin{equation}{\label{eq:a35}}
D^{[\xi]}(E,\textbf{0})=\frac{\eta^{[\xi]}}{E\!-\!\mathrm{\Delta}^{[\xi]}\!+\!h^{[\xi]}E^2\!-\!\frac{1}{5}\eta^{[\xi]}g^{[\xi]^2}J_2(E)},
\end{equation}
with
\begin{eqnarray}
J_2(E)&=&\frac{3}{2}\Bigg\{2\mu \int \frac{d^3 q}{(2 \pi )^3}\frac{[\tau_{ij}\chi _{{q}}^{(+)}(\mathbf{0})] [\tau_{ij}\chi _{{q}}^{*(+)}(\mathbf{0})]}{2\mu E -{q}^2+i\epsilon }\Bigg\}
\nonumber\\
&=&\frac{\mu}{2} \int \frac{d^3 q}{(2 \pi )^3}\frac{4q^4 +5q^2k_C^2+k_C^4}{p^2-{q}^2+i\epsilon }\frac{2 \pi \eta_q }{e^{2 \pi \eta_q}-1}\nonumber\\
&=&\frac{5}{4}k_C^2J_1(p)+(p^4-k_C^4)J_0(p)
+2\mu \int \frac{d^3 q}{(2 \pi )^3}\frac{q^4-p^4 }{p^2-{q}^2+i\epsilon }\,\frac{2 \pi \eta_q }{e^{2 \pi \eta_q}-1}\nonumber\\
&=&\frac{5}{4}k_C^2J_1(p)+(p^4-k_C^4)J_0(p)+p^2J -2\mu  \int \frac{d^3 q}{(2 \pi )^3}q^2\,\frac{2 \pi \eta_q }{e^{2 \pi \eta_q}-1}\nonumber\\
&=&W_2(p)J_0(p)-(p^2+\frac{5}{4}k_C^2)J\underbrace{-2\mu  \int \frac{d^3 q}{(2 \pi )^3}q^2\,\frac{2 \pi \eta_q }{e^{2 \pi \eta_q}-1}}_{I}.\qquad\quad\qquad\quad
\end{eqnarray}
The integral $I$ is divergent and independent of the external momentum $p$. According to the PDS regularization scheme takes the form~\cite{Ando-Shung}
\begin{eqnarray}
I=\frac{4}{3}\pi^3\mu k_C^4\Big( k_C  \zeta'(-4)-\frac{\kappa}{120}\Big),
\end{eqnarray}
with $\zeta'(-4)\approx0.00798$. If we use the cutoff regularization scheme the integral $J$ takes the form
\begin{eqnarray}\label{eq:a8}
I&\!=\!&-\frac{2\mu}{\pi}\!\int_{0}^{\mathrm{\Lambda}}\!dq q^4 \frac{\eta_q}{e^{2\pi\eta_q}-1}\nonumber\\
&\!=\!&-32\pi^3 \mu k^5_C\!\int_{\frac{2\pi k_C}{\mathrm{\Lambda}}}^{\infty}\! \frac{dx}{x^5(e^{x}-1)}\nonumber\\
&\!=\!&-32\pi^3 \mu k^5_C\Bigg\{\!\int_{0}^{\infty}\! \frac{dx}{x^5(e^{x}-1)}-\int_{0}^{\frac{2\pi k_C}{\mathrm{\Lambda}}}\frac{dx}{x^5(e^{x}-1)}\Bigg\}\! \nonumber\\
&\!=\!&-32\pi^3\mu k^5_C\Bigg\{\mathrm{\Gamma}(-4)\zeta(-4)\!-\!\int_{0}^{ \frac{2\pi k_C}{\mathrm{\Lambda}}}\!\!dx\bigg(\!\frac{1}{x^6}-\frac{1}{2x^5}+\frac{1}{12x^4}-\frac{1}{720x^2}+\mathcal{O}\,(x^0)\!\bigg)\!\!\Bigg\}\!\nonumber\\
&\!=\!&-32\pi^3\mu k^5_C\bigg\{\!\!-\frac{1}{18}\pi^2C_E \,\zeta'(-4)\! +\!\frac{1}{5}\Big(\frac{\mathrm{\Lambda}}{2\pi k_C }\Big)^{\!5}-\frac{1}{8}\Big(\frac{ \mathrm{\Lambda}}{2\pi k_C}\Big)^{\!4}+\frac{1}{36}\Big(\frac{\mathrm{\Lambda}}{2\pi k_C}\Big)^{\!3}\nonumber\\&&\,\qquad\quad\quad\,\,\,\,\,\,- \frac{1}{720}\Big(\frac{\mathrm{\Lambda}}{2\pi k_C}\Big)+\mathcal{O}\,(\frac{2\pi k_C}{\mathrm{\Lambda}})\bigg\},
\end{eqnarray}
where in the second line we use $x = 2\pi\eta_q$. Consequently, separating the integrals $J_2$ into the finite and divergent part leads to
\begin{eqnarray}{\label{eq:a38}}
J^{fin}_2&=&W_2(p)J^{fin}_0=-\frac{\mu}{2\pi}H_2(\eta_p),\\
J^{div}_2&=&W_2(p)J^{div}_0-(p^2+\frac{5}{4}k_C^2)J+I \nonumber\\
&=&p^4J^{div}_0+p^2(\frac{5}{4}k_C^2J^{div}_0\!-\!J) \!+\!(\frac{1}{4}k_C^4J^{div}_0\!-\!\frac{5}{4}k_C^2J\!+\!I).\nonumber\\
\end{eqnarray}
Thus the up-to-NLO EFT scattering amplitude for D waves is written as 
\begin{eqnarray}\label{eq:a10}
 T^{[\xi]}_{CS}=&-\frac{2\pi}{\mu} \frac{C_{0}^2(\eta_p)W_{2}(\eta_p)}{(\frac{10\pi\mathrm{\Delta}^{[\xi]}}{\eta^{[\xi]}g^{[\xi]^2}\!\mu}\!+\!\frac{2\pi}{\mu}(\frac{1}{4}k_C^4J^{div}_0\!-\!\frac{5}{4}k_C^2J\!+\!I)\!-\!\frac{1}{2}(\frac{10\pi }{\eta^{[\xi]}g^{[\xi]^2}\!\mu^2}+\frac{2\pi}{\mu}(\frac{5}{4}k_C^2J^{div}_0\!-\!J)p^2\!-\!\frac{1}{4}(\!\frac{10\pi h^{[\xi]}}{g^{[\xi]^2}\!\mu^3}\!+\frac{2\pi}{\mu}J^{div}_0)p^4\!-\!H_2(\eta_p)}.\nonumber\\
\end{eqnarray}

\noindent{\!\!The function $J_2^{div}$ has three divergences, momentum independent, momentum-squared and momentum-cubed
which are absorbed by the parameters $\mathrm{\Delta}^{[\xi]}$, $g^{[\xi]}$ and $h^{[\xi]}$
via introducing the renormalized parameters $\mathrm{\Delta}_R^{[\xi]}$, $g_R^{[\xi]}$ and $h_R^{[\xi]}$ as}
\begin{eqnarray}{\label{eq:a23}}
\mathrm{\Delta}_{\!R}^{[\xi]}&=&\frac{\mathrm{\Delta}^{[\xi]}+\frac{1}{5}\eta^{[\xi]}g^{[\xi]^2}\mu(\frac{1}{4}k_C^4J^{div}_0\!-\!\frac{5}{4}k_C^2J\!+\!I)}{1+\frac{1}{5}\eta^{[\xi]}g^{[\xi]^2}\mu(\frac{5}{4}k_C^2J^{div}_0\!-\!J)},\qquad\qquad
\end{eqnarray}
\begin{eqnarray}{\label{eq:a23}}
\frac{1}{g_R^{[\xi]^2}}&=&\frac{1}{g^{[\xi]^2}}+\frac{1}{5}\eta^{[\xi]}\mu(\frac{5}{4}k_C^2J^{div}_0\!-\!J),\\
h_{\!R}^{[\xi]}&=&\frac{h^{[\xi]}+\frac{1}{5}g^{[\xi]^2}\mu J_0^{div}}{1+\frac{1}{5}\eta^{[\xi]}g^{[\xi]^2}\mu(\frac{5}{4}k_C^2J^{div}_0\!-\!J)}.\qquad\quad\qquad\quad\quad
\end{eqnarray}
Finally, the Coulomb-subtracted EFT scattering amplitude for all possible $D$ waves are written as
\begin{eqnarray}
T^{[\xi]}_{CS}\!=\!-\frac{2\pi}{\mu}\frac{ C_0^2(\eta_p)W_2(p)}{\frac{10\pi\mathrm{\Delta}_{\!R}^{[\xi]}}{\eta^{[\xi]}g_{\!R}^{[\xi]^2}\!\mu}\!-\!\frac{1}{2}(\!\frac{10\pi }{\eta^{[\xi]}g_{\!R}^{[\xi]^2}\!\mu^2}\!)p^2\!-\!\frac{1}{4}(\!\frac{10\pi h_{\!R}^{[\xi]}}{g_{\!R}^{[\xi]^2}\!\mu^3}\!)p^4\!-\!\!H_2(\eta_p\!)}.\nonumber\\\!\!\!\!\!\!\!\!\!\!\!\!\!
\end{eqnarray}

\newpage
% BibTeX users please use
% \bibliographystyle{}
% \bibliography{}
%
% Non-BibTeX users please use

\end{document}